\documentclass[fleqn,10pt]{wlscirep}
\usepackage[utf8]{inputenc}
\usepackage[T1]{fontenc}
\usepackage{subcaption}
\usepackage{multirow}

\title{Emergence of complex structures from nonlinear interactions and noise in coevolving networks}

\author[1,2,*]{Tomasz Raducha}
\author[2,+]{Maxi San Miguel}
\affil[1]{Institute of Experimental Physics, Faculty of Physics, University of Warsaw, Pasteura 5, 02-093 Warsaw, Poland}
\affil[2]{IFISC, Institute for Cross-disciplinary Physics and Complex Systems (UIB-CSIC), Campus Universitat Illes Balears, E-07122, Palma de Mallorca, Spain}
\affil[*]{tomasz.raducha@fuw.edu.pl}
\affil[+]{maxi@ifisc.uib-csic.es}

\begin{abstract}
We study the joint effect of the non-linearity of interactions and noise
on coevolutionary dynamics. This implies a feedback loop
between the dynamics of the states of nodes and the dynamics
of the network's topology. We choose the coevolving voter
model as a prototype framework for this problem.
By numerical simulations and analytical approximations
we find three main phases
that differ in the absolute magnetization and
the size of the largest component:
a consensus phase, a coexistence phase, and
a dynamical fragmentation phase.
More detailed analysis reveals inner differences in these phases,
allowing us to divide two of them further.
In the consensus phase we can distinguish between
a weak or alternating consensus (switching
between two opposite consensus states), and
a strong consensus, in which the system
remains in the same state for the whole
realization of the stochastic dynamics.
Additionally, weak and strong consensus phases scale differently
with the system size. The strong consensus phase exists
for superlinear interactions and it is the only
consensus phase that survives in the thermodynamic limit.
In the coexistence phase we distinguish a fully-mixing
phase (both states well mixed in the network)
and a structured coexistence phase, where
the number of links connecting nodes in different
states (active links) drops
significantly due to the formation of two homogeneous communities of 
opposite states connected by a few links.
The structured coexistence phase is an example of emergence
of community structure from not exclusively topological dynamics,
but coevolution.
Our numerical observations are supported by an analytical
description using a pair approximation approach and an
ad-hoc calculation for the transition between
the coexistence and dynamical fragmentation phases.
Our work shows how simple
interaction rules including the joint effect
of non-linearity, noise, and coevolution lead to
complex structures relevant in the description of social systems.
\end{abstract}

\begin{document}

\flushbottom
\maketitle
\thispagestyle{empty}

\section*{Introduction}

Coevolving or adaptive network models \cite{gross2008adaptive} provide a better representation of 
real-world systems in comparison with static
or evolving networks .
Most empirical networks display both
network dynamics (evolution of the network's topology) as well as dynamics of the state of the nodes
\cite{albert2002statistical,kwapien2012physical}.
Moreover, a nontrivial feedback
loop between these aspects renders a simple sum
of effects analyzed separately incomplete.
Adaptive mechanisms coupling network and nodes state dynamics give rise to new phenomena absent when coevolution process is not taken into account
\cite{zimmermann2005cooperation,eguiluz2005cooperation,holme2006dynamics,raducha2018statistical,raducha2017coevolving,raducha2018predicting,gross2006epidemic,scarpino2016effect,vazquez2016rescue,fronczak2006self,toruniewska2016unstable}.
Coevolution models incorporate microscopic assumptions
in better agreement with empirical observations,
and they also produce new macroscopic results.

Another essential feature of many real-world systems
is the non-linearity associated with non-dyadic interactions.
It is often assumed in network models that
an interaction occurs pairwise, only between
two selected vertices. From a single node point
of view it means selecting one of its neighbors
at random for the interaction. This leads to
a linear relation between the number of neighbors
in a given state and the probability of choosing one of them.
However, in non-dyadic or group interactions, linearity is lost
\cite{castellano2009nonlinear}. In contagion or spreading processes, the difference between these two types of interaction goes under the name of simple vs. complex contagion \cite{centola2007cascade,centola2010spread,min2018competing}

A third crucial empirical element in many dynamical processes on networks 
is noise. This is specially important in social systems where noise is inevitable
\cite{castellano2009statistical,perc2017statistical}. It can manifest itself
on various levels. First, people chose other people 
to interact with at random. The exact form
of this randomness can take different forms, 
nevertheless the structure's evolution is never
hard-coded. But the most fundamental part of
randomness lays probably within individual choices.
For example, having exactly the same influence on two 
people's opinions we can not be sure of the outcome.
This mechanism is sometimes referred to as non-conformism \cite{sznajd2011phase}. It reflects the ability of agents to change state independently of the states of their neighbors. It is often a model parameter that needs to be calibrated to reproduce empirical data \cite{fernandez2014voter}.

In this paper we aim at exploring the joint effect of
these three important aspects  --
the coevolution of network structure and node states, 
the non-linearity of interactions and the noise --
on the behavior of the system. As the framework 
we choose the simple voter model \cite{holley1975ergodic,suchecki2005voter}.
With a binary state it is often used as a model of opinion dynamics, but different forms and extensions of the model have been fruitful in explaining empirical observations in fairly distinct phenomena such as electoral processes \cite{fernandez2014voter}, stock market \cite{carro2015markets} or online communities\cite{klimek2016dynamical}.
The consequences of coevolution \cite{vazquez2008generic,diakonova2014absorbing,toruniewska2017coupling}, noise \cite{kirman1993ants,alfarano2005estimation,carro2016noisy,peralta2018stochastic}, and non-linearity\cite{castellano2009nonlinear} have been already considered separately in the voter model.
The joint effect of these aspects, however, turns out to be more complex than a mere superposition of the results obtained so far.

The coevolving voter model (CVM) \cite{vazquez2008generic}
was among the pioneers in introducing
adaptive mechanisms in general. In the standard 
voter model, node state dynamics follows an imitation rule which is here coupled with link rewiring, introducing the coevolution. This
leads to a network fragmentation transition.
The effect of noise in the CVM \cite{diakonova2015noise} prevents the existence of absorbing configurations
so that the different phases of the system are described by dynamically active
stationary states. These include a  striking  new dynamical fragmentation phase. Additionally,
a fully-mixing phase is found, as 
could be expected for large noise levels.
Nonlinear interactions have been also considered in the CVM \cite{min2017fragmentation,raducha2018coevolving}.
Non-linearity changes
the stability of fixed points in the voter model dynamics, leading to a new
dynamically trapped coexistence phase. Finally, the joint effects of noise and non-linearity in the voter model have also been considered \cite{peralta2018analytical}. It was found that non-linearities transform a finite size transition known from the noisy voter model into a bona fide phase transition that survives in the thermodynamic limit. Here we introduce a CVM in which noise and non-linearities are jointly taken into account.

\begin{figure}
\centering
\includegraphics[scale=0.9]{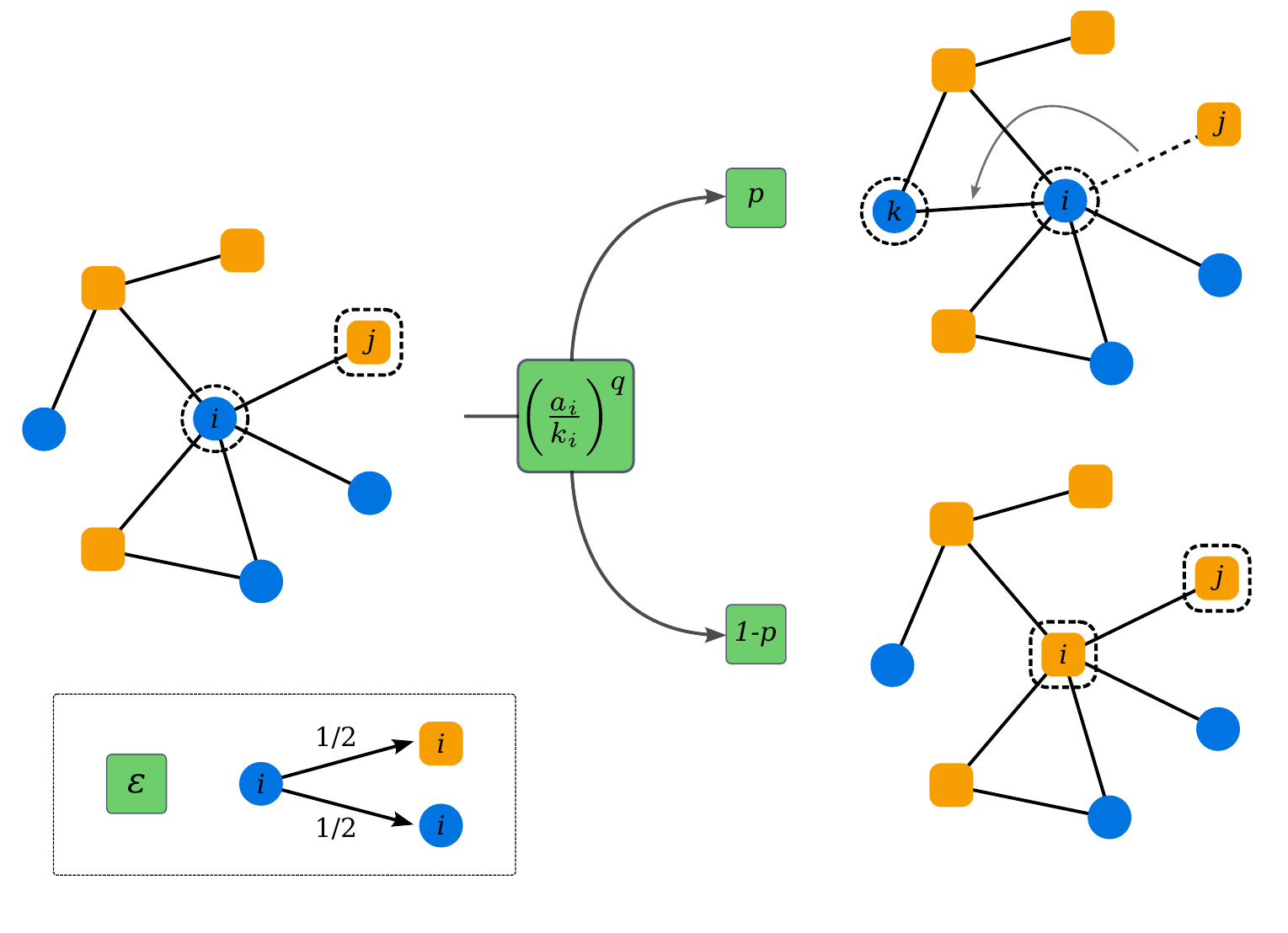}
\caption{Schematic illustration of update rules in
the nonlinear coevolving voter model with noise. Every
node is in a state $+1$ or $-1$, indicated 
by orange and blue colors. The active node $i$ is
chosen randomly. Then with probability $({a_i}/{k_i})^q$ an interaction occurs
and  one of the active links (the one to the node $j$) is selected randomly. With probability
$p$ the link $(i,j)$ is rewired to the link $(i,k)$, where $k$ is a random
node being in the same state as $i$. With
probability $1 - p$ the focal node $i$ copies the state of
the node $j$. At the end of the time step,
regardless of what happened before, the active node
draws a random state with probability~$\epsilon$.}
\label{fig:algorithm}
\end{figure}

\subsection*{The model}

Our noisy and nonlinear CVM is defined by its
time evolution over discrete time steps as follows.
First a random graph is generated and every
node is assigned a state $s_i \in \{-1,+1\}$ at 
random\footnote{Every time something is done
\textit{at random} without specifying 
the probability distribution it means the distribution
is uniform, i.e. probability is constant for every outcome.}. 
In each time step a node $i$ is chosen at random,
we call it the active or focal node. Then, with probability $({a_i}/{k_i})^q \equiv \rho_i^q$ an interaction occurs,
where $k_i$ is the degree of the focal 
node $i$, $a_i$ is the number of neighbors of the node $i$
being in the opposite state, and $q$ is 
the non-linearity parameter of the model. If an interaction occurs,
one of the $a_i$ neighbors in a different state is chosen randomly, call it~$j$.
Then, with probability $p$ a link rewiring
is performed and with complementary probability $1-p$
a state copying. When rewiring, the node $i$ 
cuts the connection to the node $j$ and creates a new link with a randomly selected node in the same
state (if there is no such node, nothing happens).
When copying the state, the node $i$ replicates 
the state of the node $j$, i.e. $s_i \to s'_i = s_j$.
At the and of the time step, regardless of
what happened before, the active node with probability
$\epsilon$ draws a random state. Note that 
this is equivalent to flipping the current state with probability
$\epsilon /2$. The algorithm of 
the model is illustrated in Figure~\ref{fig:algorithm}.
 
Our model has three parameters, namely the noise rate $\epsilon$, the plasticity $p$ and the non-linearity $q$. The parameter $p$ is a network plasticity parameter measuring the ratio of time scales of node dynamics and network dynamics. The non-linearity parameter $q$ measures the nonlinear effect of local majorities: $q=1$ corresponds to the ordinary voter model with a mechanism of random imitation, while $q<1$ indicates a probability of imitation above random imitation and $q>1$ a probability below random imitation. The ordinary voter model corresponds to $p=\epsilon=0$ and $q=1$, while $p=\epsilon=0$ corresponds to the nonlinear voter model and $p=0$ and $q=1$ to the noisy voter model. The CVM is obtained for $\epsilon=0$ and $q=1$, the noisy CVM is obtained for $q=1$ and the nonlinear CVM for $\epsilon=0$. 

Our simulations are ran from an initial random network with $N$ nodes and 
$M$ links or average degree
$\mu =  \sum_i k_i / N = 2M/N$, and with random initial conditions for the nodes states until a stationary 
state or a frozen configuration is reached (the latter being possible only for $\epsilon=0$).


\section*{Results}
We explore the space of possible values 
of the parameters ($p,q, \epsilon$) by means of computer simulations
and analytical approximations. In order 
to describe the system, we adapt typical order parameters,
such as magnetization $m = \sum_i s_i / N$, size of the largest network component $S$,
and density of active links $\rho = \sum_i a_i / 2M$.
By an active link we mean a connection
between two nodes being in opposite states. All these values are usually normalized
to fit the range $[0,1]$. Additionally, we define a new indicator,
namely community overlap ($ov.$), in order to be able to distinguish structural changes. The community overlap is a fraction of nodes assigned to the same
community by both their state and 
by the algorithm of community detection in the network\cite{clauset2004finding}.
Consequently, if a given node was assigned to the same community in both cases
it increases the $ov.$ by $1/N$ (where the denominator comes from normalization).

\begin{table}[h!]
\centering
\caption{Average values of the order parameters
in different phases. The border between phases A and B is given by $|m|=0.5$,
between phases B1 and B2  is given by $ov.=0.75$ which is approximated by
$\rho=0.1$, the border between phases B and C is given by $S=0.75$.
The difference between phases A1 and A2 can be observed in 
the dynamical behavior of the magnetization and on the system size scaling
(see Figure~\ref{fig:summary} and ~\ref{fig:mag_dist}).}
\large
\begin{tabular}{cc|c|c|c|c|}
\cline{3-6}
\multicolumn{1}{l}{}                                                                                      & \multicolumn{1}{l|}{} & \multicolumn{4}{c|}{\textbf{Phase}}                                          \\ \cline{3-6} 
\multicolumn{1}{l}{}                                                  
&      & A        & B1    & B2              & C         \\ \hline
\multicolumn{1}{|c|}{\multirow{4}{*}{\textbf{\begin{tabular}[c]{@{}c@{}}Order\\ parameter\end{tabular}}}}
& ~~$\langle |m| \rangle$~~ & ~~$\lesssim1$~~ & ~~$\gtrsim0$~~   & ~~$\gtrsim0$~~  & ~~$\gtrsim0$ ~~        \\
\cline{2-6} 
\multicolumn{1}{|c|}{}     
& ~~$\langle S \rangle$~~       & ~~$1$~~      & ~~$1$~~ & ~~$\lesssim1$~~ & ~~$\gtrsim0.5$~~ \\ \cline{2-6} 
\multicolumn{1}{|c|}{}                                                                  
& ~~$\langle ov. \rangle$~~    & ~~~--~~~   & ~~$\gtrsim0.5$~~  & ~~$\lesssim1$~~ & ~~$1$~~    \\ \cline{2-6} 
\multicolumn{1}{|c|}{}                                                                 
& ~~$\langle \rho \rangle$~~   & ~~~$\gtrsim0$~~~   & finite &  ~~$<0.1$~~   & ~~$\gtrsim0$~~    \\ \hline
\end{tabular}
\label{tab:params}
\end{table}

\subsection*{Phase diagram}

\begin{figure}
\centering
\includegraphics[scale=0.85]{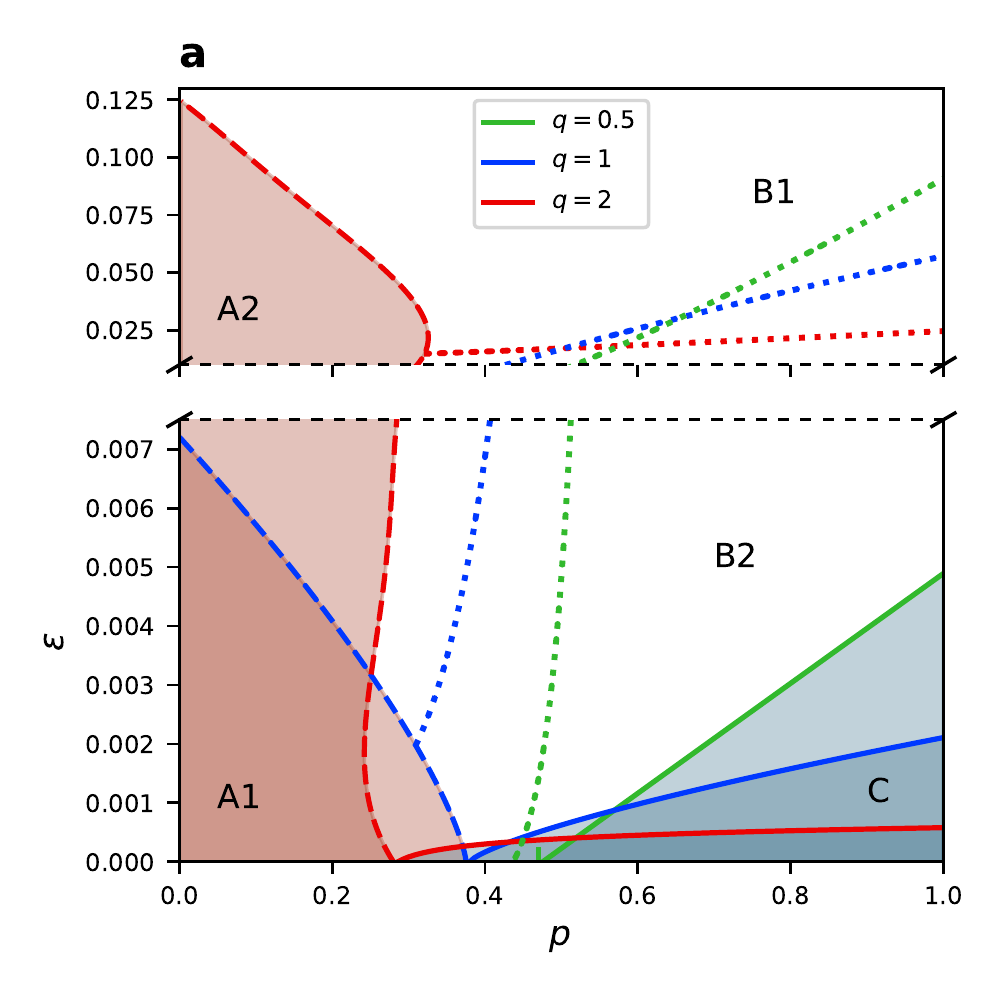}
\includegraphics[scale=0.85]{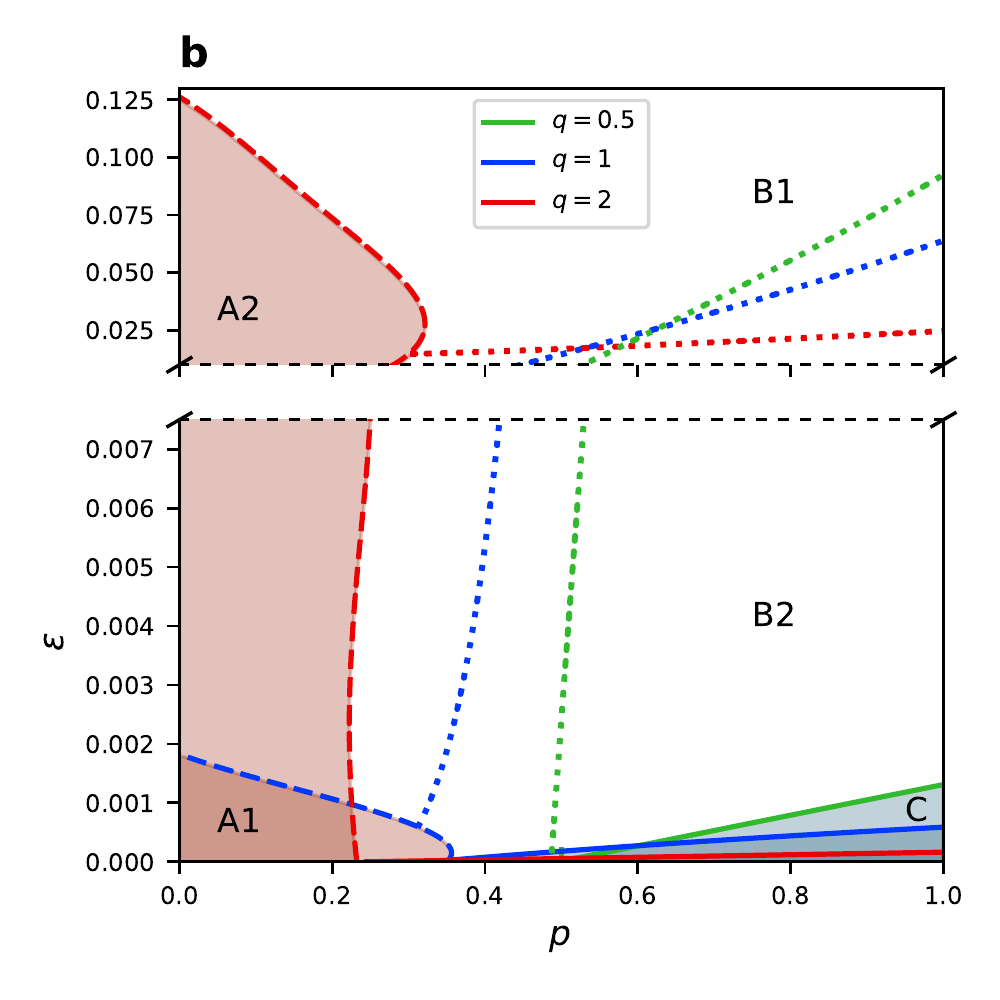}
\caption{Phase diagram in the $p$-$\epsilon$ space for (a) $N=250$ and (b) $N=1000$
for $\mu=4$ and different values of $q$ (indicated by color). Results
based on simulations
averaged over 500 realizations.
The red area represents the phase A, the white one phase B,
and the blue one phase C. The border between
phases A and B is a line defined by the medium value of the average (over time in one stochastic realization) absolute 
magnetization $\langle |m| \rangle=0.5$ (dashed lines). The border between
phases B and C is a line defined by the medium size of 
the largest component $\langle S \rangle=0.75$ (solid lines). 
The border between phases B1 and B2 (dotted lines) is approximated by $\langle \rho \rangle=0.1$ (see Table~\ref{tab:params}).}
\label{fig:summary}
\end{figure}

We numerically study the $p$-$\epsilon$ phase diagram for three different values
of the $q$ parameter -- the sublinear case $q=0.5$,
the ordinary linear case $q=1$, and the superlinear
case $q=2$. These phase diagrams are shown in
Figure~\ref{fig:summary} for two different network sizes. Obviously,
for any finite amount of noise in the system
a frozen configuration does not exist, and any phase
is described by a characteristic dynamical stationary state.
We can distinguish three general phases 
in the model. Phase A, indicated by the red area in the figure,
is a consensus phase. In this range of parameters 
the system stays in a consensus state for most of the time,
i.e. magnetization is close to $\pm 1$ and 
the network is connected having a single large component $S=1$ and a small number of active links $\rho\gtrsim0$ . 
If we increase the noise rate $\epsilon$ or the plasticity $p$
sufficiently, we obtain phase B indicated by the white area in  Figure~\ref{fig:summary}, and referred to 
as a fully-mixing phase in previous work for the noisy CVM \cite{diakonova2015noise}.
In this phase the magnetization drops 
to zero, $m=0$, hence there is no consensus in the system any more. In addition, the network stays connected most of the time, $S\lesssim1$ . We refer to this phase as a
coexistence phase. As we will see, phase B
is not homogeneous in its whole range of parameters and it can be either fully-mixing (phase B1) or structured (phase B2), what is indicated by different values of the density of active links $\rho$ and community overlap $ov.$
Finally, for  values of the rewiring
probability above the critical point $p_c$ of the nonlinear
CVM \cite{min2017fragmentation},
and relatively small noise rates, phase C arises.
It is marked by the blue area in the figure.
In this region we find dynamical fragmentation --
the network consists of two separate components
with nodes in each one being in opposite states, so that $m\gtrsim0$, $S\gtrsim0.5$ and $\rho=0$. It is possible, however, that the two network components 
get connected intermittently in the stationary state due to
noise and random rewiring, creating again a single component network with $m\gtrsim0$ and a small number of active links $\rho\gtrsim0$.
Phase C can be described as dynamical switching between these two arrangements. 
Values of all analyzed quantities for every phase
are summarized in Table~\ref{tab:params}.

\begin{figure}
\centering
\includegraphics[scale=0.85]{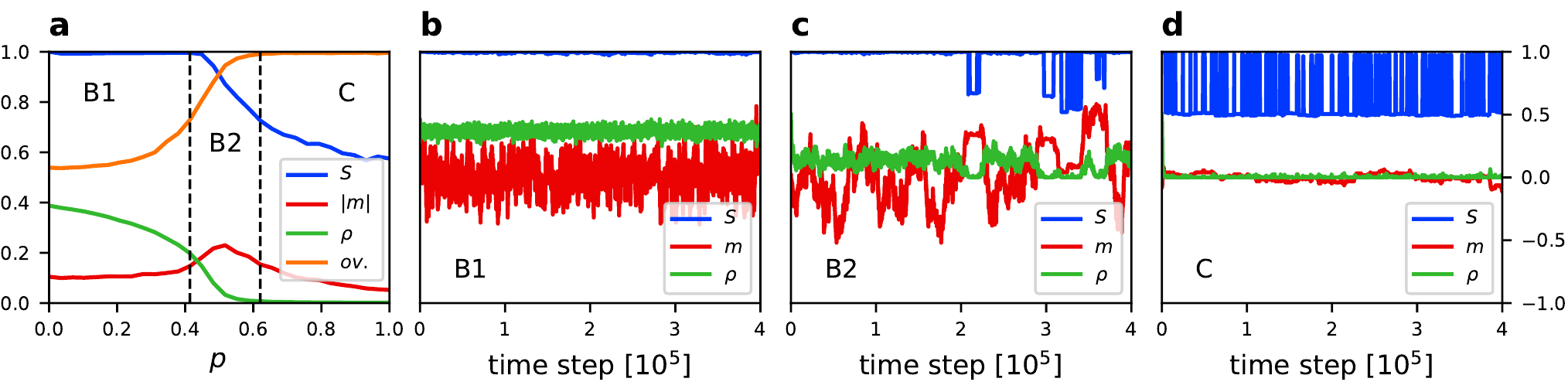}
\caption{(a) Average value of the size of the largest component $\langle S \rangle$, absolute magnetization $\langle|m| \rangle$,
density of active links $\langle \rho \rangle$, and community overlap $\langle ov. \rangle$ vs. rewiring probability $p$
for $q=0.5$, $\epsilon=0.001$, $N=250$, and $\mu=4$.
Results averaged over 500 simulation runs.
Borders between phases are indicated by dashed
lines according to the Table~\ref{tab:params}.
For every phase a trajectory of $S$, $m$ and $\rho$ is given for the same
parameters values and (b) $p=0.1$, (c) $p=0.46$, (d) $p=0.8$.
}
\label{fig:cross_q05}
\end{figure}

\begin{figure}
\centering
\includegraphics[scale=0.85]{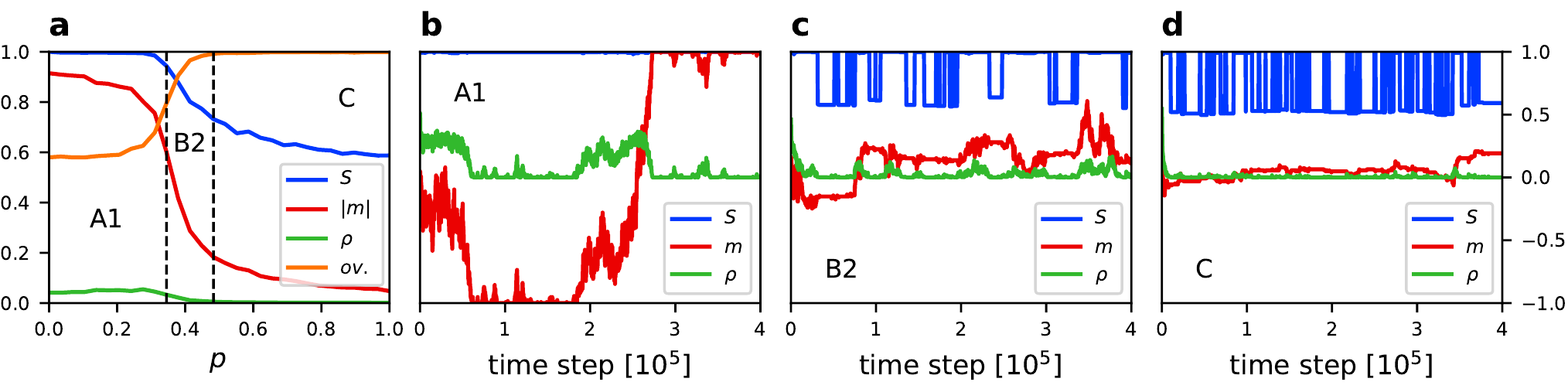}
\caption{(a) Average value of the size of the largest component $\langle S \rangle$, absolute magnetization $\langle|m| \rangle$,
density of active links $\langle \rho \rangle$, and community overlap $\langle ov. \rangle$  vs. rewiring probability $p$
for $q=1$, $\epsilon=0.0005$, $N=250$, and $\mu=4$.
Results averaged over 500 simulation runs.
Borders between phases are indicated by dashed
lines according to Table~\ref{tab:params}.
For every phase a trajectory of $S$, $m$ and $\rho$ is given for the same
parameters values and (b) $p=0.1$, (c) $p=0.38$, (d) $p=0.6$.}
\label{fig:cross_q1}
\end{figure}

\begin{figure}
\centering
\includegraphics[scale=0.85]{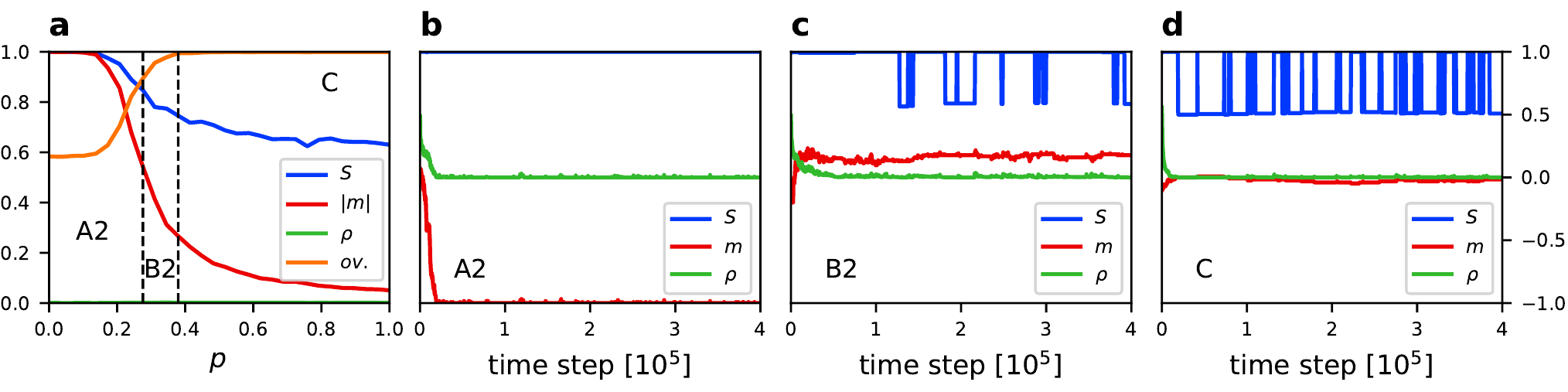}
\caption{(a)  Average value of the size of the largest component $\langle S \rangle$, absolute magnetization $\langle|m| \rangle$,
density of active links $\langle \rho \rangle$, and community overlap $\langle ov. \rangle$  vs. rewiring probability $p$
for $q=2$, $\epsilon=0.0002$, $N=250$, and $\mu=4$.
Results averaged over 500 simulation runs.
Borders between phases are indicated by dashed
lines according to Table~\ref{tab:params}.
For every phase a trajectory of $S$, $m$ and $\rho$ is given for the same
parameters values and (b) $p=0.2$, (c) $p=0.32$, (d) $p=0.9$.}
\label{fig:cross_q2_down}
\end{figure}

\begin{figure}
\centering
\includegraphics[scale=0.85]{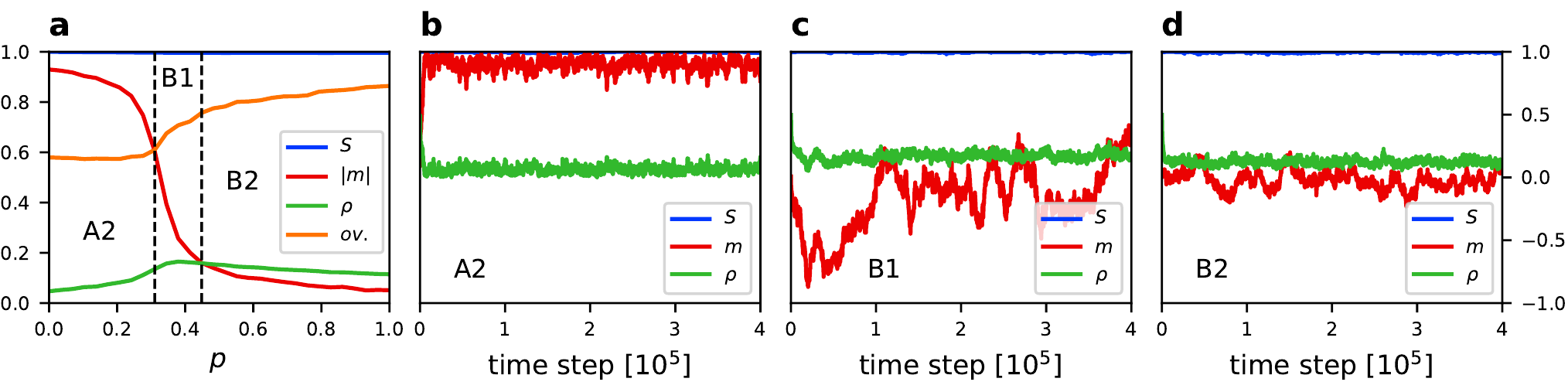}
\caption{(a)  Average value of the size of the largest component $\langle S \rangle$, absolute magnetization $\langle|m| \rangle$,
density of active links $\langle \rho \rangle$, and community overlap $\langle ov. \rangle$  vs. rewiring probability $p$
for $q=2$, $\epsilon=0.03$, $N=250$, and $\mu=4$.
Results averaged over 500 simulation runs.
Borders between phases are indicated by dashed
lines according to Table~\ref{tab:params}.
For every phase a trajectory of $S$, $m$ and $\rho$ is given for the same
parameters values and (b) $p=0.1$, (c) $p=0.38$, (d) $p=0.9$.}
\label{fig:cross_q2_up}
\end{figure}

For the linear case ($q=1$) phases A and C exist only for a finite size of the network,
and the size of these phases in the parameter space decreases with growing number of nodes.
For
the sublinear scenario $q<1$, we can see in 
Figure~\ref{fig:summary} that the same holds for phase C, while 
phase A does not exist at all.
The only point where the average absolute magnetization slightly raises
is at $p_c$ and for $\epsilon \approx 0$, but its maximal value is only about 0.3.
This raise is due to higher fluctuations close to the transition point.
On the other hand,  phase C
prevails for even twice larger noise rate than in the linear scenario.

In the superlinear case ($q=2$)  phase C
is much smaller in parameter space and disappears faster with growing system size. But phase A
prevails for much larger noise than in the linear case $q=1$. We  observe
phase A even for $\epsilon$ larger by almost
two orders of magnitude. Additionally, the system size scaling  is different for the
superlinear scenario. Indeed, non-linearity has significant influence
on the nature of phase A. For $q<1$ it 
does not exist, for $q=1$ it exists only in finite networks,
and for $q>1$ phase A persists in the thermodynamic limit.

\subsection*{A closer look at the phases}

In order to better understand the behavior of the system and the differences between phases
we analyse horizontal cross-sections of Figure~\ref{fig:summary} and single-run trajectories which are
presented in Figures~\ref{fig:cross_q05}-\ref{fig:cross_q2_up} for the linear, sublinear and superlinear cases.
In panel (a) of each of them, a phase diagram with respect to rewiring probability $p$
and for a particular value of $\epsilon$ is presented.
Values of the noise rate are chosen
in such a way that allows to show three phases in one panel. For an horizontal
cross-section of the full phase diagram it is difficult to capture all phases.
Therefore, areas of the middle phase can be narrow, but still different
values of the order parameters can be distinguished. Panels (b)-(d) show typical time traces of the order parameters in different phases.

For the sublinear case ($q=0.5$) we can see the differences between phase B1(fully-mixing, Figure~\ref{fig:cross_q05}b) and phase B2 (structured, Figure~\ref{fig:cross_q05}c). 
Phases B1 and B2 have zero average absolute magnetization
$\langle |m| \rangle \approx 0$, but we can distinguish a region with high
density of active links (B1) and small density of active links (B2). Phase C
also has a low density of active links, but the largest network component switches from $S=1$ to $S=0.5$, giving and average value 
$\langle S \rangle \approx 0.5$, whereas phase B2 on the average stays connected.

Results for the linear case ($q=1$) are shown in Figure~\ref{fig:cross_q1}. Phase A is characterized by a magnetization which tends to stay at one of the consensus states, but it
can switch from $-1$ to $+1$, or the other way around, during the time evolution.
Therefore, $\langle |m| \rangle \approx 1$ but $\langle m \rangle =0$.
To distinguish it from the superlinear case where $\langle m \rangle \approx \pm 1$ we call this phase A1.
We also observe that for $q=1$ in phase B2 (Figure~\ref{fig:cross_q1}c) the network can fragment close to the transition line, however
it remains a single component network most of the time. 

Figure~\ref{fig:cross_q2_down} and Figure~\ref{fig:cross_q2_up} correspond, respectively, to small and large noise rates in the superlinear case ($q=2$).
In this scenario the consensus phase A prevails for much larger noise rate.
In panels (b) of  Figure~\ref{fig:cross_q2_down} and Figure~\ref{fig:cross_q2_up}
we can see how the system behaves in phase A for $q=2$. It quickly reaches a consensus
state for either $m=1$ or $m=-1$  and remains at this value of magnetization. Therefore, $\langle |m| \rangle \approx 1$ and
in contrast to the linear case in a time average also $\langle m \rangle \approx \pm 1$. To account for this
difference we call the consensus phase A2 for $q>1$ .

\begin{figure}
\centering
\includegraphics[scale=0.85]{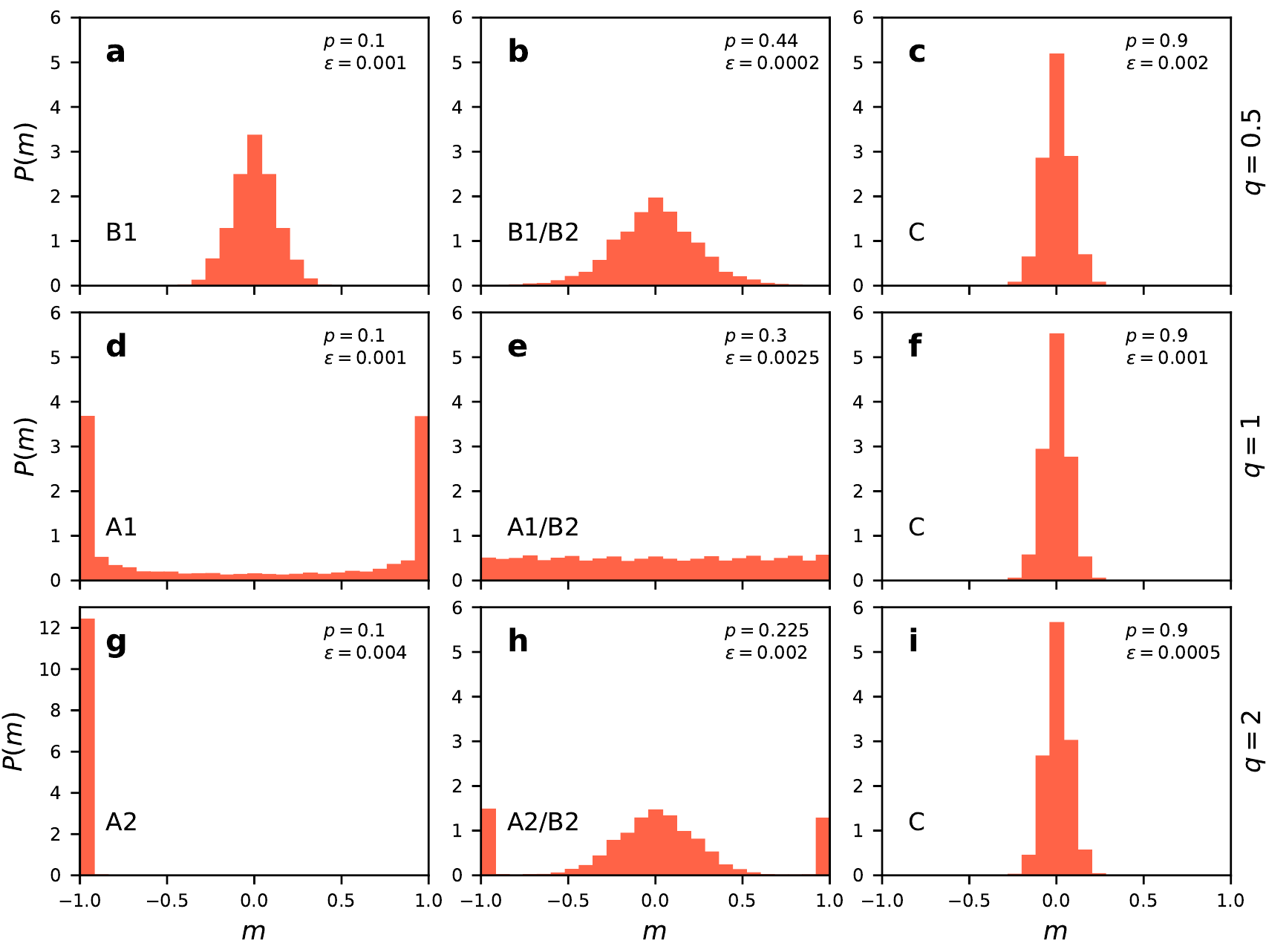}
\caption{Probability distribution of 
the magnetization $m$ for $N=250$ and $\mu=4$ averaged over
$10^7$ MC steps after thermalization. Results for (a, b, c) $q=0.5$,
(d, e, f) $q=1$, and
(g, h, j) $q=2$.
Values of the rewiring probability $p$, noise rate $\epsilon$
and the phase indication
are given in the panels.
}
\label{fig:mag_dist}
\end{figure}

The difference between the consensus phase A1 and A2 is also clearly visible from the probability distribution of the 
magnetization in a given realization of the dynamical process (Figure~\ref{fig:mag_dist}).
In the linear case there is a bimodal distribution for the magnetization
with two equal peaks at values $+1$
and $-1$ (Figure~\ref{fig:mag_dist}d), while for the superlinear case there is
a single peak for a value of the magnetization at either of the
boundary values $+1$
or $-1$, depending on the run (Figure~\ref{fig:mag_dist}g).
For $q=2$, once the consensus is reached the system stays there with minor fluctuations (phase A2), while
for $q = 1$ the system goes back and forth between opposite consensus states (phase A1).
Furthermore, phase A2 is robust against finite-size fluctuations,
while phase A1 disappears in the thermodynamic limit \cite{diakonova2015noise}
(see Figure~\ref{fig:summary}).

The distribution of the magnetization gives additional insights on the phase diagram: The fact that phase A does not exist in the sublinear case ($q < 1$), is reflected in a distribution with  a single peak at $0$ for all values of $\epsilon$ (Figure~\ref{fig:mag_dist}a-c). However, the variance of the distribution takes its maximal value for noise going to zero
and $p=p_c$, i.e. close to the transition 
point between coexistence and fragmentation phase
in the nonlinear CVM \cite{min2017fragmentation}. A different form of the transition between phases A and B for the linear and superlinear case is also observed. For $q=1$ there is a flat distribution at the transition point (Figure~\ref{fig:mag_dist}e) , while a trimodal distribution is found for $q=2$ 
(Figure~\ref{fig:mag_dist}h). 
A trimodal magnetization distribution was reported
before in the noisy voter model on a static network \cite{peralta2018analytical},
but only for non-linearity parameter equal 5 or larger.
With  coevolution, trimodality here is obtained already for $q=2$.

\begin{figure}
\centering
\includegraphics[scale=0.85]{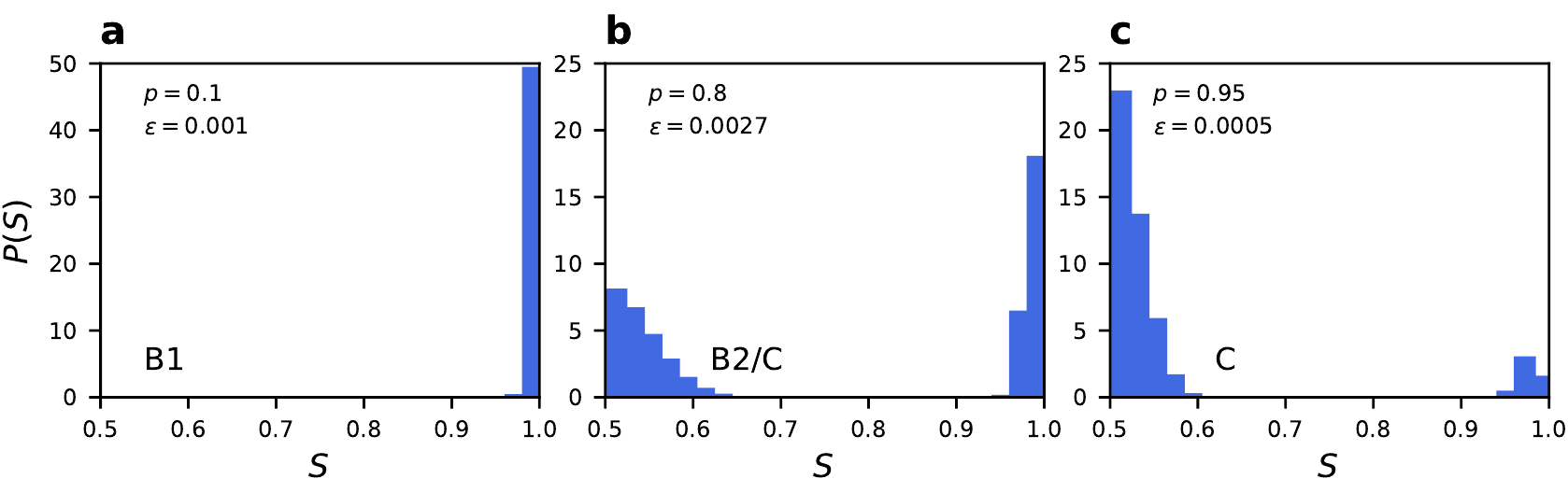}
\caption{Probability distribution of the size of the largest component $S$
for $N=250$, $\mu=4$, and $q=0.5$ averaged over
$10^7$ MC steps after thermalization. Results for (a)  phase B1, (b) close
to the transition line and (c)  phase C.
Values of the rewiring probability $p$ and noise rate $\epsilon$ are given in
the panels.}
\label{fig:comp_dist}
\end{figure}

Phase C can be defined in terms of the size of the largest component.
In phases A and B it is
equal to the size of the whole network ($S=1$),
while the phase C is characterized by a dynamical fragmentation
into two components of similar size and opposite state. Due to noise expressed in random
changes of nodes states and rewiring the components
are constantly being reconnected and disconnected.
It can be examined looking at the trajectory or at
the probability distribution of the size of the largest component, which is presented in  Figure~\ref{fig:comp_dist}.

\subsection*{Community structure}

\begin{figure}
\centering
\includegraphics[scale=0.9]{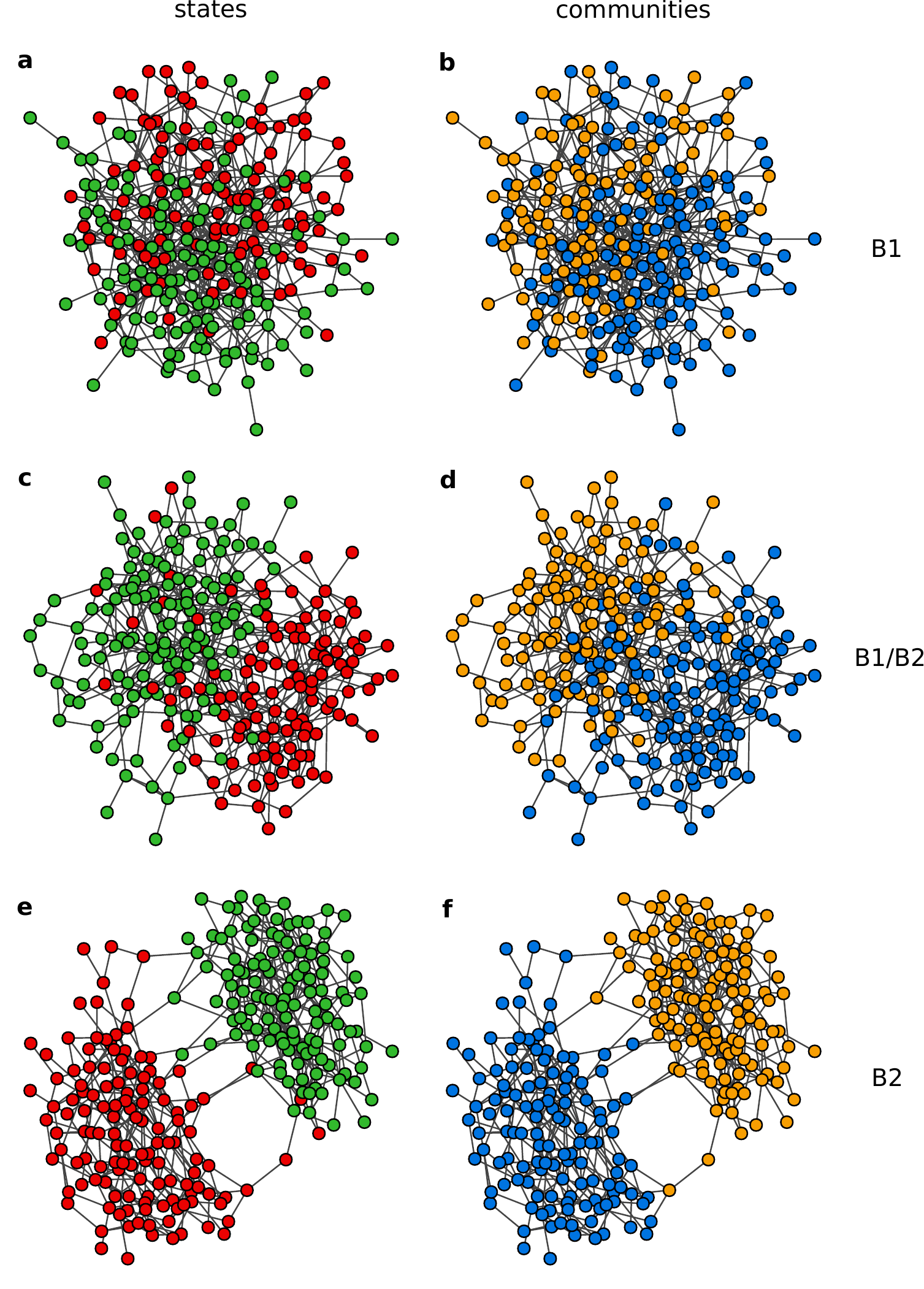}
\caption{Examples of the network topology and node states in the stationary state for
(a, b) phase B1 at $p=0.9$, $\epsilon=0.004$, (c, d) the transition line
between B1 and B2 at $p=0.5$, $\epsilon=0.025$, and (e, f)  phase B2 at
$p=0.5$, $\epsilon=0.1$. In all cases  $N=250$, $M=500$
 and $q=2$.
State $+1$ is indicated by green color and $-1$ by red (a, c, e).
Communities found by the community detection algorithm \cite{clauset2004finding}
are colored blue and orange (b, d, f).
In the  B1 communities are not associated with the states of the nodes,
so that the overlap between (a) and (b) is 0.54. On the transition line states
start to rearrange
into communities giving overlap equal 0.77 for (c) and (d).
In phase B2 the communities defined by the state of the nodes 
are well overlapping with the structural communities giving overlap of 0.97
for (e) and (f).}
\label{fig:b_phase}
\end{figure}

Phase B is generally defined by zero average
magnetization, also zero average absolute magnetization,
and by the existence of one large network component.
Nonetheless, this description leaves room for different possible configurations.
Analysis of the trajectories showed that
the density of active links can vary within phase B,
but the question is weather this is a sign of a topological change.
In the linear case ($q=1$) only the fully-mixing phase was reported \cite{diakonova2015noise},
with nodes of states $+1$ and $-1$
well mixed inside a random graph. We refer to this configuration as phase B1.
On the other hand,
we can satisfy conditions for the phase B having two evident
communities, highly connected internally and
of opposite states, with only a few links bridging them.
There is still zero magnetization and
one large network component in such configuration  characterized by a small number of active links. We call this phase B2.
The difference between
phases B1 and B2 is clearly seen in Figure~\ref{fig:b_phase}.

Although the difference between phases B1 and B2 can be seen in
the density of active links, 
a closer look at Figure~\ref{fig:b_phase} suggests that phase B2 has well
defined topological communities.
Therefore,  we propose an alternative quantitative measure for the difference between phases B1 and B2, the overlap between state communities --
defined by the state of the nodes -- and
structural communities found by a community detection algorithm. We use a classical algorithm from
\cite{clauset2004finding}, but the result 
does not differ much when using other algorithms.
Each node is assigned to the state community by its state and to a structural community
by the algorithm's result. The relative overlap between these
two communities is a new quantitative indicator of the phase of the system\footnote{Note that we have
to take the maximum overlap from two possible
community assignments. If we have structural communities
\textit{a} and \textit{b}, we can associate
community \textit{a} with the state $+1$ and community
\textit{b} with $-1$, or the other way around.
Therefore, in a perfect overlap with wrong assignment
one can get zero overlap. Trying both
possibilities and taking maximum solves this issue.}.
For a random assignment or no community structure
the overlap will be close to $0.5$.
This is the situation in phase B1.
For  phase B2 the overlap
will be close to 1.  This means that our dynamical coevolving model generates clear
topological communities emerging from local interactions
involving only state of nodes. This result may potentially explain
process of formation of communities in social networks,
where such structures are especially common\cite{girvan2002community}.

\subsection*{Identifying phase boundaries}

So far, we gave a description of different phases with different qualitative behavior. Transition lines between these different types of behavior are not clearly or unambiguously defined because every analyzed phase indicator changes value significantly  across the phase diagram.
We do not focus on properties of
phase transitions in this work, but rather on the properties of
different emerging structures.
Therefore we follow here a simple and pragmatic
way to identify boundaries between different phases,
employing previous approaches\cite{diakonova2015noise}. These boundaries have to be understood as an arbitrary way to deal with crossover system behavior.

Each of the order parameters or phase indicators analyzed has a continuous range of values (for a large $N$) and different phases are
described by the extreme values of these quantities,
allowed in this range. Therefore, a straightforward way of dividing
the phase diagram into separate phases is to use
the middle value, i.e. the value in the middle between
the maximum and minimum of a given range. For example, the absolute magnetization
is defined in the range $[0,1]$, taking a value close to 1 in phase A and a value close to zero in phase B. Hence, we identify the 
border between the two phases with $\langle |m| \rangle=0.5$. 

Likewise, the size of the normalized largest network component takes values in the range $[0.5,1]$. In phases A and B there is a single component network ($S=1$),  while phase C is characterized by the dynamical fragmentation
into two components of similar size and opposite state, so that $S=0.5$. In this phase, due to noise expressed in random
changes of nodes states and rewiring, the components
are constantly being reconnected and disconnected. We then identify  \cite{diakonova2015noise} the border between phase B and phase C by the middle value $S=0.75$. 
This is a line at which 
the network is half of the time fragmented and half of the time
contains only one big component. This phase boundary can also be obtained from
the probability distribution of the size of 
the largest component 
(Figure~\ref{fig:comp_dist}). At the transition line (panel b) two peaks have the same area.

Finally the boundary between phases B1 and B2 can be identified in terms of the the overlap ($ov.$), which takes values in the range $[0.5, 1]$.
There is a transition from values around $0.5$ in B1, up to $1$ in B2, and so we define the transition line at $ov.=0.75$.  However, this parameter is computationally
very demanding. Alternatively, we can approximate the identification of the transition line by a small value of the density of active links which we arbitrarily fix at $\rho=0.1$

\subsection*{Analytical predictions}

\begin{figure}
\centering
\includegraphics[scale=0.85]{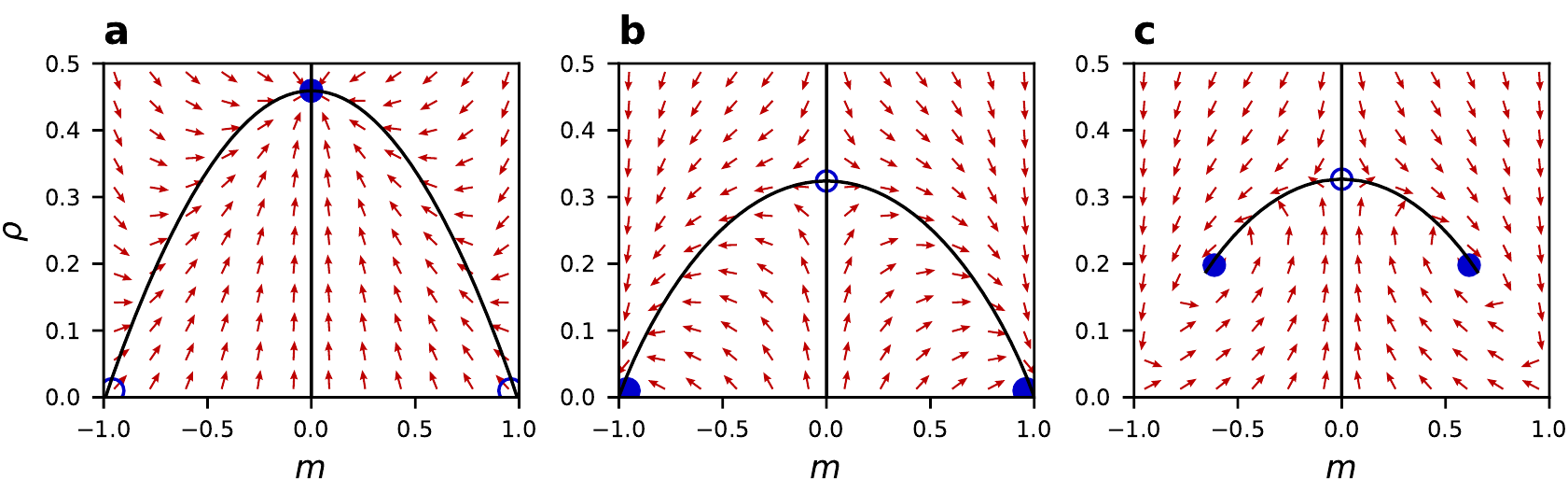}
\caption{Flow diagram of the system dynamics in 
the $m$-$\rho$ space for $\mu=8$, $p=0.1$ and
(a) $q=0.5$, $\epsilon=0$, (b) $q=2$, $\epsilon=0$, (c) $q=2$, $\epsilon=0.1$.
Arrows represent the dynamical
direction of the system according to the pair approximation dynamics
(Equations~\ref{eqn:pair_approximation}). 
Fixed points are represented by full circles (stable)
and empty circles (unstable). Note how non-linearity 
and noise can change the stability and position of the fixed points.} 
\label{fig:flow_diagram}
\end{figure}

\begin{figure}
\centering
\includegraphics[scale=0.85]{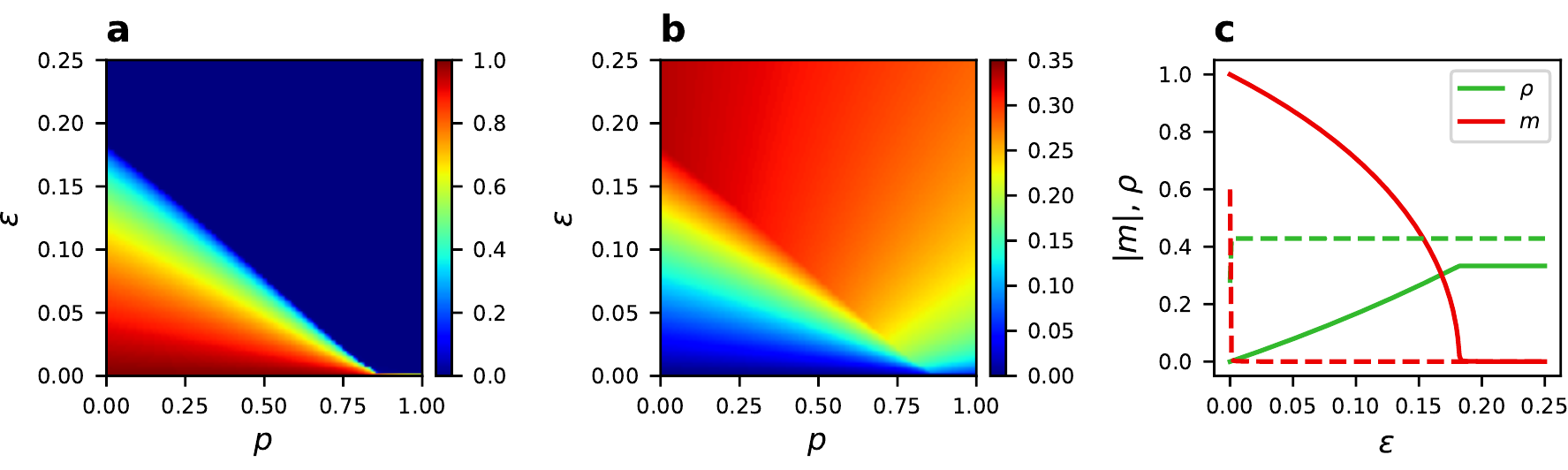}
\caption{Numerical solution of the pair approximation 
(Equations~\ref{eqn:pair_approximation}). (a) Absolute magnetization $|m|$ and (b)
density of active links $\rho$ for $q=2$ and $\mu=8$ in the
$p$-$\epsilon$ phase diagram
showing the existence of  phase A2 in the thermodynamic limit.
(c) Absolute magnetization and density of active links vs. noise rate for the static
case ($p=0$) and $\mu=8$.
A continuous transition is present for $q=2$ (solid lines),
in contrast to the case of $q=1$ (dashed lines), where
consensus can only be obtained  for $\epsilon=0$.}
\label{fig:q2_solutions}
\end{figure}

The magnetization $m$ and the density of active links $\rho$ 
obey Equations~\ref{eqn:pair_approximation}
describing the dynamics of the system, as
derived in the Methods section. Several fixed points $(m^*,\rho^*)$
of these ordinary differential equations
can be found depending on parameter values.
However, not all of them are stable, therefore
not all of them are observed in numerical simulations.
To analyze the stability of these fixed points,
we consider flow diagrams of the dynamics in  Figure~\ref{fig:flow_diagram}.
Note, that from panel (a) to (b) only the value of $q$ changes, emphasizing the difference between sublinear and superlinear cases, while
from  panel (b) to (c) only the value of $\epsilon$ changes, emphasizing the noise effect in the superlinear case.
A change in the non-linearity parameter $q$ can reverse
the stability of fixed points when going over the boundary
value of 1. A change in the noise rate $\epsilon$
can additionally shift the position of fixed points
allowing for fixed points different than $m=-1,0,1$.
Since the analytical description is derived
in the thermodynamic limit, we don't observe stable
fixed points at non-zero magnetization for $q \leq 1$.
This finding is consistent with the scaling
behavior of numerical results indicating existence
only of phase B in the limit of a large number of nodes $N$.
For the superlinear case, although the fixed
points are placed at the same values of the magnetization
$m=-1,0,1$ (for $p<p_c$, $\epsilon<\epsilon_c$),
their stability is inverted -- now only the solutions of $|m|=1$
are stable, corresponding to phase A2. This is clearly seen in the analytical
prediction for the phase diagram in 
Figure~\ref{fig:q2_solutions}.
These results, obtained in the thermodynamic limit, indicate that phase A2 should be
observed for any $N$ when $q=2$, which 
is in agreement with our numerical results
in  Figure~\ref{fig:summary}.
Separate mean-field prediction of the disappearance
of phase A1 in the thermodynamic limit
was given by Diakonova et. al. \cite{diakonova2015noise}
for the linear CVM with noise. Additionally, phase C is not obtained in the thermodynamic limit.

The non-linear noisy voter model in a fixed network ($p=0$) has been thoroughly
analytically studied \cite{peralta2018analytical}, showing that
 $q=1$ is a bordering value between a unimodal and bimodal distribution of
the magnetization $m$. In other words, it is 
a transition line between existence and nonexistence of phase A.
The agreement of our results with previous studies can be seen
when analyzing the extreme value of $p=0$ in the phase diagrams
(Figure~\ref{fig:q2_solutions}a and b). It is also separately presented in the
Figure~\ref{fig:q2_solutions}c. The transition to phase A, characterized by nonzero absolute magnetization $|m|$, exists for finite values
of $\epsilon$ only when $q>1$. In the Methods section we derive a formula
for the critical value of the noise rate $\epsilon_{c}(p=0)$ at which
the system looses the consensus state, that is, the transition form phase A to phase B (Equation~\ref{eqn:eps_p_eq_0}).
This result is in agreement with the numerical solution of
Equations~\ref{eqn:pair_approximation} presented
in Figure~\ref{fig:q2_solutions}c, giving a critical value of the density of active links
$\rho_c=\frac{1}{3} \approx0.33$ and $\epsilon_{c}=\frac{2}{11}\approx 0.18$
for the parameters values used in the figure. The analytical solution
from Equation~\ref{eqn:eps_p_eq_0} also predicts disappearance 
of the transition at $q=1$.
In the reference\cite{peralta2018analytical} a similar prediction of the 
critical noise rate was given for a complete graph:
$\epsilon'_c(p=0) = 2^{-q}(q-1)$ which would give for the case of
Figure~\ref{fig:q2_solutions}c $\epsilon'_c = 0.25$.
Therefore, a complete graph gives only a first approximation to the value found here.

\begin{figure}
\centering
\includegraphics[scale=0.85]{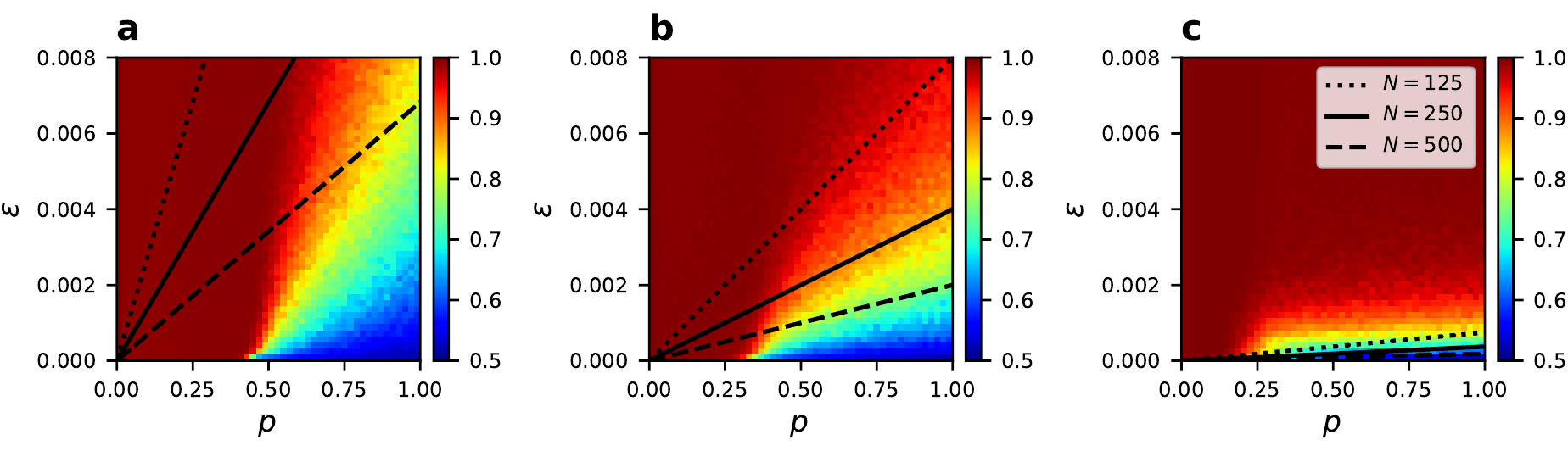}
\caption{Analytical prediction of the transition line between phases B and C,
according to Equation~\ref{eqn:phase_c}, compared with simulation results
for $N=250$, $\mu=4$, and (a) $q=0.5$, (b) $q=1$, (c) $q=2$. The analytical 
prediction is indicated by a solid black line.
Scaling with the network size (dotted and dashed lines)
and dependence on the non-linearity parameter $q$
give trends consistent with simulations.
Numerical results are averaged over 500 realizations.}
\label{fig:phase_c}
\end{figure}

There has been no previous attempt of an analytical approximation describing the transition from phase B to the
dynamical fragmentation phase C already found in the noisy linear CVM. We propose here a simple
description of this phenomena.
Having two separate,
but internally homogeneous network components (clusters)  with nodes in opposite states, the only
way of connecting them is by 
a random change of node's state and  link rewiring
to the second component. The probability of
the first event is independent of $q$ and is simply
given by $\epsilon /2$. When the node changing state
is selected as the active node the probability
of an interaction is $\rho_i^q$.
Since $\rho_i \in [0,1]$, for smaller $q$
the probability of an interaction is higher,
except for boundary cases with $\rho_i=0,1$. 
To reconnect the two clusters, rewiring must occur,
but this happens always with probability $p$, despite
the value of $q$. Therefore, for a single node
in a state opposite to the  whole cluster, the
probability of connecting to the other cluster
is constant (since $\rho_i=1$). However,
once the two clusters are connected, the higher
probability of an interaction for lower $q$
means a higher probability of rewiring causing fragmentation again.
Consequently, we expect phase C to persist
for larger noise when $q$ is smaller. More detailed
description of this process is given in the Methods
section, where we derive the following  formula for
the transition line:
\begin{equation}\label{eqn:phase_c}
\epsilon_S (p) = \frac{4}{N} \left( \frac{1}{\mu} \right)^{2q} (2^q + 2) p .
\end{equation}
Based on this approximation
we predict phase C to fade with growing
non-linearity parameter $q$ or with growing system size $N$,
as shown in Figure~\ref{fig:phase_c}.
Both predictions are consistent with
our numerical results.

\section*{Discussion}

In this paper we analyzed the nonlinear coevolving voter model with noise.
Depending on the values of the three main parameters
-- the rewiring probability $p$,
the noise intensity $\epsilon$, and the non-linearity
parameter $q$ -- we observed
three distinct phases: a consensus phase A,
a coexistence phase B, and a dynamical fragmentation phase C. We observed, however,
significant internal differences within phases A and B.
The first one can be further divided into
phase A1 and A2. Phase A1, for $q = 1$,
is a consensus phase with absolute magnetization
equal 1 on average, but real magnetization
switching between $-1$ and $+1$ states, giving rise to a
bimodal magnetization distribution within one realization of the stochastic dynamics.
In phase A2, observed for $q>1$, there is a stable
consensus, i.e. global magnetization states $-1$ and $+1$ are stable.
Consequently, during one realization of the stochastic dynamics the system
remains in a given consensus state, producing a unimodal magnetization
distribution with a peak at the maximal (minimal)
magnetization. Additionally, phases A1 and A2 have different system size scaling 
-- phase A1  disappears in the thermodynamic limit, while phase A2
is stable against finite size fluctuations. Finally, phase A does not exist for $q<1$.

Phase B can be similarly
divided into phases B1 and B2. Phase B1 is a 
fully-mixing phase with random network
structure and random states of the nodes, giving zero magnetization. 
But for larger $p$ and low noise intensity we observe phase B2, which
has the same vanishing average magnetization,
but a different network structure. In the structured phase B2 one can easily
distinguish two communities of opposite
states connected by just a few inter-community links.
This structural difference is confirmed by community detection algorithms.

Phase C is associated with a dynamical fragmentation
of the network -- two components in opposite magnetization
states are being constantly connected and disconnected.
We derive an analytical description of this behavior and an approximated value for the transition line between phase B2 and phase C.

The only phases surviving in the thermodynamic
limit are phases A2, B1 and B2. The transition line form phase A2 to phase B is largely independent of finite size fluctuations. 
We have also presented an analytical pair approximation able to describe these findings and the main features of phases A and B in the thermodynamic limit.

Our work fills a gap in the studies of the CVM.
It provides a binding bridge between
studies on the CVM with noise \cite{diakonova2015noise}
and studies on the nonlinear
CVM \cite{min2017fragmentation}. It reduces to the nonlinear noisy voter model
\cite{peralta2018analytical} and the ordinary
CVM \cite{vazquez2008generic} for a proper
configuration of parameters values. We obtain
full consistency with those limiting cases
and we explore new parameter domains.
Our work brings the analysis of the voter model
to a greater complexity by taking into account
the joint effect of noise, coevolution and non-linearity which turns out not to be a mere superposition of them. It may provide
a tool for the evaluation of the relevance of different
mechanisms in the description of opinion dynamics,
but can be also a reference point in the study of coevolving network models.
We also show how nonlinear vs linear interactions can change the
stability of a consensus state in the network and
how topological communities can arise from non-topological interactions.
These results are of relevant value in the description of
social networks.


\section*{Methods}

\subsection*{Pair approximation}

We use the same approach as used for the nonlinear coevolving voter model \cite{min2017fragmentation}
to describe the dynamics of magnetization $m$ and the density of active links $\rho$.
Given the network homogeneity due 
to the random rewiring, we assume each node
to have the same average degree $\mu = 2M/N$. 
Let us denote by $n_+ = (1+m) /2$ and $n_- =(1-m)/2$
the fraction of nodes in the state $+1$ and $-1$ respectively. Then, when we pick a node
in the state $\pm 1$ as the active node, 
the probability of choosing a neighbor in the opposite state
is given by ${\rho}/{2n_{\pm}}$. In other words, $\frac{\rho}{2n_{\pm}}$
gives the density of active links $\rho_i$ for 
a node $i$ being in the state $s_i = \pm 1$.
Therefore, the probability of an interaction 
is given by $\rho_i^q = ({\rho}/{2n_{\pm}})^q \equiv n_q^{\pm}$. When $q$ takes integer values this can be also interpreted as 
the probability of 
choosing a neighbor in the opposite state $q$ times.
Hence, when an interaction occurs with this probability,
we can make the approximation that there is at least $q$ neighbors 
in the opposite state and for the rest of them
the probability of being in a different state than 
the focal node is ${\rho}/{2n_{\pm}}$. All together 
this implies that\footnote{This is a rough estimate, 
more precise one could be obtained using Bayes' theorem,
but it doesn't display a significant difference in the results.}
$a_i \approx q + (\mu - q) \frac{\rho}{2n_{\pm}}$.

To approximate the evolution of the density of active links $\rho$ we must estimate the
contributions of different events that can result in a change of $\rho$.
These events are: (i) rewiring, followed by a change of state through noise,
(ii) rewiring, without a change of the node's state due to noise, (iii) changing the state of
the node through state copying with no further change due
to noise, and (iv) changing the state of the node only as a result of
noise, with no previous state copying or rewiring.
Let $\delta_{\pm}$ be the change in the total number of active links
given that a node $i$ such that $s_i =\pm 1$ 
changed state. The total change in the number
of active links in the four possible events above is: (i) $1 + \delta_{\pm}$, (ii) $-1$,
and for events (iii) and (iv) just $\delta_{\pm}$. 
Magnetization changes only when the state of the node
is changed via copying or as a result of noise in three possible scenarios: 
state copying with no noise effect,
link rewiring followed by noise effect, and no interaction
-- neither state copying nor link rewiring --
but noise acting alone. When the focal node 
having a state $s_i =\pm 1$ changes its state, the total change
in the magnetization is $\mp 2$.
Hence, in the thermodynamic limit we have:
\begin{equation}\label{eqn:pair_approx_long}
\begin{split}
 \frac{dm}{dt} = & 2 (1-p)(1-\frac{\epsilon}{2}) (n_- n_q^- - n_+ n_q^+) +
2 p \frac{\epsilon}{2} (n_- n_q^- - n_+ n_q^+) +
2 \frac{\epsilon}{2}  \left[n_- (1- n_q^-) - n_+ (1- n_q^+)\right] ,
\\
 \frac{d\rho}{dt} = & \frac{2}{\mu} \bigg \{
  p \frac{\epsilon}{2}  \left[n_+ n_q^+ (1+\delta_+) + n_- n_q^- (1+\delta_-)\right] -
  p (1-\frac{\epsilon}{2}) (n_+ n_q^+ + n_- n_q^- ) +
  \\
&  + (1-p) (1-\frac{\epsilon}{2}) (n_+ n_q^+ \delta_+ + n_- n_q^- \delta_-) +
  \frac{\epsilon}{2} \left[n_+ (1- n_q^+) \delta_+ + n_- (1- n_q^-) \delta_- \right]
\bigg \} ,
\end{split}
\end{equation}
With a few simple algebraic transformations these equations can be rewritten as:
\begin{equation}\label{eqn:pair_approximation}
\begin{split}
& \frac{dm}{dt} = 2 (1-p)(1-\epsilon) (n_- n_q^- - n_+ n_q^+) + \epsilon (n_- - n_+) ,
\\
& \frac{d\rho}{dt} = \frac{2}{\mu} \left[
 (1-p)(1-\epsilon) (n_+ n_q^+ \delta_+ + n_- n_q^- \delta_-)
 - p(n_+ n_q^+ + n_- n_q^- ) 
+ \frac{\epsilon}{2} (n_+  \delta_+ + n_-  \delta_-)
\right] .
\end{split}
\end{equation}
When node $i$ changes state,
all its $a_i$ active links 
become inactive and all other $\mu - a_i$
inactive links become active, therefore the total
change in the number of active links is
$\delta_{\pm} = \mu - 2a_i$. Using the previous approximation
for $a_i$ we can write
$\delta_{\pm} = \mu - 2q - 2(\mu - q) \frac{\rho}{2n_{\pm}}$.

The simplest stationary solution (fixed point) of Equations~\ref{eqn:pair_approximation} is given
by taking the magnetization $m=0$,
which leads to an equation for the stationary value of $\rho$:
\begin{equation}\label{eqn:stat_eq}
- \rho^{q+1} 2 (\mu -q) (1-p) (1-\epsilon)
+ \rho^q  [(1-p)(1-\epsilon)(\mu-2q) -p ]
- \rho \epsilon (\mu - q)
+ \frac{\epsilon}{2} (\mu - 2q)
= 0 .
\end{equation}
A fixed point solution with $m = \pm 1$ does not exist for any finite noise rate $\epsilon$, whilst
for $\epsilon=0$ a stationary solution is $\rho = 0$.
Setting the noise rate to zero together with
$m=0$ we obtain the stationary solution
of the nonlinear CVM \cite{min2017fragmentation}:
\begin{equation}\label{eqn:stat_ncvm}
\rho^*(\epsilon=0) = \frac{(1-p)(\mu - 2q) - p}{2 (1-p)(\mu -q)} .
\end{equation}
For $\epsilon=0$ and $q=1$ we recover the solution of the
standard CVM \cite{vazquez2008generic}:
\begin{equation}\label{eqn:stat_cvm}
\rho^*(\epsilon=0,q=1) = \frac{(1-p)(\mu -1) - 1}{2 (1-p)(\mu -1)} .
\end{equation}

To compare our results with the nonlinear noisy voter model
on static networks we analyze our approximation for the particular case $p=0$.
Putting $\rho = \rho_c (1-m^2)$ in the first of Equations~\ref{eqn:pair_approximation} and
performing a stability analysis of the fixed point 
solution $m=0$ we can find a critical noise value
\begin{equation}\label{eqn:eps_p_eq_0}
\epsilon_{c}(p=0) = \frac{2 \rho_c^q (q-1)}{2 \rho_c^q (q-1) + 1} ,
\end{equation}
which depends on the critical value of the density of active links $\rho_c$.
The latter can be obtained from
 Equation~\ref{eqn:stat_eq} as $\rho_c = \frac{\mu - 2q}{2\mu - 2q}$.
This formula is shown to be in  full agreement with numerical solutions
of Equations~\ref{eqn:pair_approximation} shown in 
Figure~\ref{fig:q2_solutions}c.

\subsection*{Phase C: Finite size scaling}
In order to describe the behavior of the dynamical fragmentation
phase we first look for an approximation for the
probabilities of reconnecting two separate clusters
and of disconnecting two clusters sharing
at most two links. In this approach we omit events
of probability proportional to $(1/N)^3$
and to $\epsilon^2$, or of higher order.

Imagine two separate and internally homogeneous
components of opposite states, as it happens
in phase C. The simplest way of connecting
them under the rules of the nonlinear noisy CVM 
involves two steps. First, one of the nodes,
call it $i$, must change it's state. This is possible only due to noise
and occurs with probability  $\epsilon/2$.
Second, node $i$ that has changed its state
must rewire one of its links to a node in the opposite
cluster, having the same state.
This can happen with probability  $p \rho_i^q / N$,
because we need to select this particular node
as the active node ($1/N$), an interaction
has to occur ($\rho_i^q$), and rewiring must be performed ($p$).
Note, that since node $i$ is the only node
in a different state than its cluster $\rho_i^q = 1$.
Finally, it gives the probability of reconnecting two components equal:
\begin{equation}\label{eqn:P_r}
P_r = \frac{\epsilon}{2} \frac{p}{N}.
\end{equation}

Approaching the transition line between
phases B and C now from phase B, so that a fragmentation event occurs, we consider
one single component network disconnecting into two equal
clusters. As done before, imagine a situation two
time steps before a possible fragmentation --
network has two internally homogeneous components in opposite.
One of the nodes $i$ is part of a bridge, i.e. it is connected
to two nodes in the opposite cluster. Now,
for the fragmentation to occur we need both of the links
between the components to be rewired. 
The probability to rewire the first one is
$\frac{1}{N}(2 / \mu)^q p (1-\frac{\epsilon}{2})+\frac{2}{N} (1 / \mu)^q p$.
We have to select the node $i$ ($1/N$) or one
of its neighbors ($2/N$). An interaction must occur,
what happens with probability $(a_j / \mu)^q$,
where the number of active links is 2 for  node $i$
and 1 for each of its neighbors in the opposite cluster. Finally, a rewiring
must occur with probability $p$. Additionally
if node $i$ was selected, it can not change
its state due to noise ($1 - \frac{\epsilon}{2}$),
otherwise fragmentation could not be achieved
in two steps. The transition occurs, however,
for very small values of noise and therefore
we can approximate $1 - \frac{\epsilon}{2} \approx 1$.
To rewire the second link we have to select one
of the two nodes ($2/N$) connecting the link, an interaction must occur
$(1 / \mu^q)$, which must be a rewiring event ($p$).
Therefore, the probability of losing the last link
between the two clusters is $\frac{2}{N} (1 / \mu)^q p $.
Finally, we obtain the probability of disconnecting
two clusters sharing only two links:
\begin{equation}\label{eqn:P_d}
P_d = \left[ \frac{1}{N} (2 / \mu)^q p + \frac{2}{N} (1 / \mu)^q p \right] \frac{2}{N} (1 / \mu)^q p.
\end{equation}

Between phases B and C a continuous fragmenting
and reconnecting of the network is observed.
We define the transition between the two phases when connection and fragmentation happens at such a rate that half of the time
the system consists of two separate components
and half of the time the network is connected.
Therefore, at the transition line we expect
$P_r = P_d$, which leads to the equation for the critical density of noise given in the main text (Equation~\ref{eqn:phase_c}).


\bibliography{bibliography}

\begin{thebibliography}{10}
\urlstyle{rm}
\expandafter\ifx\csname url\endcsname\relax
  \def\url#1{\texttt{#1}}\fi
\expandafter\ifx\csname urlprefix\endcsname\relax\def\urlprefix{URL }\fi
\expandafter\ifx\csname doiprefix\endcsname\relax\def\doiprefix{DOI: }\fi
\providecommand{\bibinfo}[2]{#2}
\providecommand{\eprint}[2][]{\url{#2}}

\bibitem{gross2008adaptive}
\bibinfo{author}{Gross, T.} \& \bibinfo{author}{Blasius, B.}
\newblock \bibinfo{journal}{\bibinfo{title}{Adaptive coevolutionary networks: a
  review}}.
\newblock {\emph{\JournalTitle{Journal of the Royal Society Interface}}}
  \textbf{\bibinfo{volume}{5}}, \bibinfo{pages}{259--271}
  (\bibinfo{year}{2008}).

\bibitem{albert2002statistical}
\bibinfo{author}{Albert, R.} \& \bibinfo{author}{Barab{\'a}si, A.-L.}
\newblock \bibinfo{journal}{\bibinfo{title}{Statistical mechanics of complex
  networks}}.
\newblock {\emph{\JournalTitle{Reviews of modern physics}}}
  \textbf{\bibinfo{volume}{74}}, \bibinfo{pages}{47} (\bibinfo{year}{2002}).

\bibitem{kwapien2012physical}
\bibinfo{author}{Kwapie{\'n}, J.} \& \bibinfo{author}{Dro{\.z}d{\.z}, S.}
\newblock \bibinfo{journal}{\bibinfo{title}{Physical approach to complex
  systems}}.
\newblock {\emph{\JournalTitle{Physics Reports}}}
  \textbf{\bibinfo{volume}{515}}, \bibinfo{pages}{115--226}
  (\bibinfo{year}{2012}).

\bibitem{zimmermann2005cooperation}
\bibinfo{author}{Zimmermann, M.~G.} \& \bibinfo{author}{Egu{\'\i}luz, V.~M.}
\newblock \bibinfo{journal}{\bibinfo{title}{Cooperation, social networks, and
  the emergence of leadership in a prisoner’s dilemma with adaptive local
  interactions}}.
\newblock {\emph{\JournalTitle{Physical Review E}}}
  \textbf{\bibinfo{volume}{72}}, \bibinfo{pages}{056118}
  (\bibinfo{year}{2005}).

\bibitem{eguiluz2005cooperation}
\bibinfo{author}{Egu{\'\i}luz, V.~M.}, \bibinfo{author}{Zimmermann, M.~G.},
  \bibinfo{author}{Cela-Conde, C.~J.} \& \bibinfo{author}{Miguel, M.~S.}
\newblock \bibinfo{journal}{\bibinfo{title}{Cooperation and the emergence of
  role differentiation in the dynamics of social networks}}.
\newblock {\emph{\JournalTitle{American journal of sociology}}}
  \textbf{\bibinfo{volume}{110}}, \bibinfo{pages}{977--1008}
  (\bibinfo{year}{2005}).

\bibitem{holme2006dynamics}
\bibinfo{author}{Holme, P.} \& \bibinfo{author}{Ghoshal, G.}
\newblock \bibinfo{journal}{\bibinfo{title}{Dynamics of networking agents
  competing for high centrality and low degree}}.
\newblock {\emph{\JournalTitle{Physical review letters}}}
  \textbf{\bibinfo{volume}{96}}, \bibinfo{pages}{098701}
  (\bibinfo{year}{2006}).

\bibitem{raducha2018statistical}
\bibinfo{author}{Raducha, T.}, \bibinfo{author}{Wili\'nski, M.},
  \bibinfo{author}{Gubiec, T.} \& \bibinfo{author}{Stanley, H.~E.}
\newblock \bibinfo{journal}{\bibinfo{title}{Statistical mechanics of a
  coevolving spin system}}.
\newblock {\emph{\JournalTitle{Physical Review E}}}
  \textbf{\bibinfo{volume}{98}}, \bibinfo{pages}{030301}
  (\bibinfo{year}{2018}).

\bibitem{raducha2017coevolving}
\bibinfo{author}{Raducha, T.} \& \bibinfo{author}{Gubiec, T.}
\newblock \bibinfo{journal}{\bibinfo{title}{Coevolving complex networks in the
  model of social interactions}}.
\newblock {\emph{\JournalTitle{Physica A: Statistical Mechanics and its
  Applications}}} \textbf{\bibinfo{volume}{471}}, \bibinfo{pages}{427--435}
  (\bibinfo{year}{2017}).

\bibitem{raducha2018predicting}
\bibinfo{author}{Raducha, T.} \& \bibinfo{author}{Gubiec, T.}
\newblock \bibinfo{journal}{\bibinfo{title}{Predicting language diversity with
  complex networks}}.
\newblock {\emph{\JournalTitle{PloS one}}} \textbf{\bibinfo{volume}{13}},
  \bibinfo{pages}{e0196593} (\bibinfo{year}{2018}).

\bibitem{gross2006epidemic}
\bibinfo{author}{Gross, T.}, \bibinfo{author}{D’Lima, C. J.~D.} \&
  \bibinfo{author}{Blasius, B.}
\newblock \bibinfo{journal}{\bibinfo{title}{Epidemic dynamics on an adaptive
  network}}.
\newblock {\emph{\JournalTitle{Physical review letters}}}
  \textbf{\bibinfo{volume}{96}}, \bibinfo{pages}{208701}
  (\bibinfo{year}{2006}).

\bibitem{scarpino2016effect}
\bibinfo{author}{Scarpino, S.~V.}, \bibinfo{author}{Allard, A.} \&
  \bibinfo{author}{H{\'e}bert-Dufresne, L.}
\newblock \bibinfo{journal}{\bibinfo{title}{The effect of a prudent adaptive
  behaviour on disease transmission}}.
\newblock {\emph{\JournalTitle{Nature Physics}}} \textbf{\bibinfo{volume}{12}},
  \bibinfo{pages}{1042--1046} (\bibinfo{year}{2016}).

\bibitem{vazquez2016rescue}
\bibinfo{author}{Vazquez, F.}, \bibinfo{author}{Serrano, M.~{\'A}.} \&
  \bibinfo{author}{San~Miguel, M.}
\newblock \bibinfo{journal}{\bibinfo{title}{Rescue of endemic states in
  interconnected networks with adaptive coupling}}.
\newblock {\emph{\JournalTitle{Scientific reports}}}
  \textbf{\bibinfo{volume}{6}}, \bibinfo{pages}{29342} (\bibinfo{year}{2016}).

\bibitem{fronczak2006self}
\bibinfo{author}{Fronczak, P.}, \bibinfo{author}{Fronczak, A.} \&
  \bibinfo{author}{Ho{\l}yst, J.~A.}
\newblock \bibinfo{journal}{\bibinfo{title}{Self-organized criticality and
  coevolution of network structure and dynamics}}.
\newblock {\emph{\JournalTitle{Physical Review E}}}
  \textbf{\bibinfo{volume}{73}}, \bibinfo{pages}{046117}
  (\bibinfo{year}{2006}).

\bibitem{toruniewska2016unstable}
\bibinfo{author}{Toruniewska, J.}, \bibinfo{author}{Suchecki, K.} \&
  \bibinfo{author}{Ho{\l}yst, J.~A.}
\newblock \bibinfo{journal}{\bibinfo{title}{Unstable network fragmentation in
  co-evolution of potts spins and system topology}}.
\newblock {\emph{\JournalTitle{Physica A: Statistical Mechanics and its
  Applications}}} \textbf{\bibinfo{volume}{460}}, \bibinfo{pages}{1--15}
  (\bibinfo{year}{2016}).

\bibitem{castellano2009nonlinear}
\bibinfo{author}{Castellano, C.}, \bibinfo{author}{Mu{\~n}oz, M.~A.} \&
  \bibinfo{author}{Pastor-Satorras, R.}
\newblock \bibinfo{journal}{\bibinfo{title}{Nonlinear q-voter model}}.
\newblock {\emph{\JournalTitle{Physical Review E}}}
  \textbf{\bibinfo{volume}{80}}, \bibinfo{pages}{041129}
  (\bibinfo{year}{2009}).

\bibitem{centola2007cascade}
\bibinfo{author}{Centola, D.}, \bibinfo{author}{Egu{\'\i}luz, V.~M.} \&
  \bibinfo{author}{Macy, M.~W.}
\newblock \bibinfo{journal}{\bibinfo{title}{Cascade dynamics of complex
  propagation}}.
\newblock {\emph{\JournalTitle{Physica A: Statistical Mechanics and its
  Applications}}} \textbf{\bibinfo{volume}{374}}, \bibinfo{pages}{449--456}
  (\bibinfo{year}{2007}).

\bibitem{centola2010spread}
\bibinfo{author}{Centola, D.}
\newblock \bibinfo{journal}{\bibinfo{title}{The spread of behavior in an online
  social network experiment}}.
\newblock {\emph{\JournalTitle{science}}} \textbf{\bibinfo{volume}{329}},
  \bibinfo{pages}{1194--1197} (\bibinfo{year}{2010}).

\bibitem{min2018competing}
\bibinfo{author}{Min, B.} \& \bibinfo{author}{San~Miguel, M.}
\newblock \bibinfo{journal}{\bibinfo{title}{Competing contagion processes:
  Complex contagion triggered by simple contagion}}.
\newblock {\emph{\JournalTitle{Scientific reports}}}
  \textbf{\bibinfo{volume}{8}}, \bibinfo{pages}{1--8} (\bibinfo{year}{2018}).

\bibitem{castellano2009statistical}
\bibinfo{author}{Castellano, C.}, \bibinfo{author}{Fortunato, S.} \&
  \bibinfo{author}{Loreto, V.}
\newblock \bibinfo{journal}{\bibinfo{title}{Statistical physics of social
  dynamics}}.
\newblock {\emph{\JournalTitle{Reviews of modern physics}}}
  \textbf{\bibinfo{volume}{81}}, \bibinfo{pages}{591} (\bibinfo{year}{2009}).

\bibitem{perc2017statistical}
\bibinfo{author}{Perc, M.} \emph{et~al.}
\newblock \bibinfo{journal}{\bibinfo{title}{Statistical physics of human
  cooperation}}.
\newblock {\emph{\JournalTitle{Physics Reports}}}  (\bibinfo{year}{2017}).

\bibitem{sznajd2011phase}
\bibinfo{author}{Sznajd-Weron, K.}, \bibinfo{author}{Tabiszewski, M.} \&
  \bibinfo{author}{Timpanaro, A.~M.}
\newblock \bibinfo{journal}{\bibinfo{title}{Phase transition in the sznajd
  model with independence}}.
\newblock {\emph{\JournalTitle{EPL (Europhysics Letters)}}}
  \textbf{\bibinfo{volume}{96}}, \bibinfo{pages}{48002} (\bibinfo{year}{2011}).

\bibitem{fernandez2014voter}
\bibinfo{author}{Fern{\'a}ndez-Gracia, J.}, \bibinfo{author}{Suchecki, K.},
  \bibinfo{author}{Ramasco, J.~J.}, \bibinfo{author}{San~Miguel, M.} \&
  \bibinfo{author}{Egu{\'\i}luz, V.~M.}
\newblock \bibinfo{journal}{\bibinfo{title}{Is the voter model a model for
  voters?}}
\newblock {\emph{\JournalTitle{Physical review letters}}}
  \textbf{\bibinfo{volume}{112}}, \bibinfo{pages}{158701}
  (\bibinfo{year}{2014}).

\bibitem{holley1975ergodic}
\bibinfo{author}{Holley, R.~A.} \& \bibinfo{author}{Liggett, T.~M.}
\newblock \bibinfo{journal}{\bibinfo{title}{Ergodic theorems for weakly
  interacting infinite systems and the voter model}}.
\newblock {\emph{\JournalTitle{The annals of probability}}}
  \bibinfo{pages}{643--663} (\bibinfo{year}{1975}).

\bibitem{suchecki2005voter}
\bibinfo{author}{Suchecki, K.}, \bibinfo{author}{Egu{\'\i}luz, V.~M.} \&
  \bibinfo{author}{San~Miguel, M.}
\newblock \bibinfo{journal}{\bibinfo{title}{Voter model dynamics in complex
  networks: Role of dimensionality, disorder, and degree distribution}}.
\newblock {\emph{\JournalTitle{Physical Review E}}}
  \textbf{\bibinfo{volume}{72}}, \bibinfo{pages}{036132}
  (\bibinfo{year}{2005}).

\bibitem{carro2015markets}
\bibinfo{author}{Carro, A.}, \bibinfo{author}{Toral, R.} \&
  \bibinfo{author}{San~Miguel, M.}
\newblock \bibinfo{journal}{\bibinfo{title}{Markets, herding and response to
  external information}}.
\newblock {\emph{\JournalTitle{PloS one}}} \textbf{\bibinfo{volume}{10}},
  \bibinfo{pages}{e0133287} (\bibinfo{year}{2015}).

\bibitem{klimek2016dynamical}
\bibinfo{author}{Klimek, P.}, \bibinfo{author}{Diakonova, M.},
  \bibinfo{author}{Egu{\'\i}luz, V.~M.}, \bibinfo{author}{San~Miguel, M.} \&
  \bibinfo{author}{Thurner, S.}
\newblock \bibinfo{journal}{\bibinfo{title}{Dynamical origins of the community
  structure of an online multi-layer society}}.
\newblock {\emph{\JournalTitle{New Journal of Physics}}}
  \textbf{\bibinfo{volume}{18}}, \bibinfo{pages}{083045}
  (\bibinfo{year}{2016}).

\bibitem{vazquez2008generic}
\bibinfo{author}{Vazquez, F.}, \bibinfo{author}{Egu{\'\i}luz, V.~M.} \&
  \bibinfo{author}{San~Miguel, M.}
\newblock \bibinfo{journal}{\bibinfo{title}{Generic absorbing transition in
  coevolution dynamics}}.
\newblock {\emph{\JournalTitle{Physical review letters}}}
  \textbf{\bibinfo{volume}{100}}, \bibinfo{pages}{108702}
  (\bibinfo{year}{2008}).

\bibitem{diakonova2014absorbing}
\bibinfo{author}{Diakonova, M.}, \bibinfo{author}{San~Miguel, M.} \&
  \bibinfo{author}{Egu{\'\i}luz, V.~M.}
\newblock \bibinfo{journal}{\bibinfo{title}{Absorbing and shattered
  fragmentation transitions in multilayer coevolution}}.
\newblock {\emph{\JournalTitle{Physical Review E}}}
  \textbf{\bibinfo{volume}{89}}, \bibinfo{pages}{062818}
  (\bibinfo{year}{2014}).

\bibitem{toruniewska2017coupling}
\bibinfo{author}{Toruniewska, J.}, \bibinfo{author}{Ku{\l}akowski, K.},
  \bibinfo{author}{Suchecki, K.} \& \bibinfo{author}{Ho{\l}yst, J.~A.}
\newblock \bibinfo{journal}{\bibinfo{title}{Coupling of link-and node-ordering
  in the coevolving voter model}}.
\newblock {\emph{\JournalTitle{Physical Review E}}}
  \textbf{\bibinfo{volume}{96}}, \bibinfo{pages}{042306}
  (\bibinfo{year}{2017}).

\bibitem{kirman1993ants}
\bibinfo{author}{Kirman, A.}
\newblock \bibinfo{journal}{\bibinfo{title}{Ants, rationality, and
  recruitment}}.
\newblock {\emph{\JournalTitle{The Quarterly Journal of Economics}}}
  \textbf{\bibinfo{volume}{108}}, \bibinfo{pages}{137--156}
  (\bibinfo{year}{1993}).

\bibitem{alfarano2005estimation}
\bibinfo{author}{Alfarano, S.}, \bibinfo{author}{Lux, T.} \&
  \bibinfo{author}{Wagner, F.}
\newblock \bibinfo{journal}{\bibinfo{title}{Estimation of agent-based models:
  the case of an asymmetric herding model}}.
\newblock {\emph{\JournalTitle{Computational Economics}}}
  \textbf{\bibinfo{volume}{26}}, \bibinfo{pages}{19--49}
  (\bibinfo{year}{2005}).

\bibitem{carro2016noisy}
\bibinfo{author}{Carro, A.}, \bibinfo{author}{Toral, R.} \&
  \bibinfo{author}{San~Miguel, M.}
\newblock \bibinfo{journal}{\bibinfo{title}{The noisy voter model on complex
  networks}}.
\newblock {\emph{\JournalTitle{Scientific reports}}}
  \textbf{\bibinfo{volume}{6}}, \bibinfo{pages}{24775} (\bibinfo{year}{2016}).

\bibitem{peralta2018stochastic}
\bibinfo{author}{Peralta, A.}, \bibinfo{author}{Carro, A.},
  \bibinfo{author}{San~Miguel, M.} \& \bibinfo{author}{Toral, R.}
\newblock \bibinfo{journal}{\bibinfo{title}{Stochastic pair approximation
  treatment of the noisy voter model}}.
\newblock {\emph{\JournalTitle{New Journal of Physics}}}
  \textbf{\bibinfo{volume}{20}}, \bibinfo{pages}{103045}
  (\bibinfo{year}{2018}).

\bibitem{diakonova2015noise}
\bibinfo{author}{Diakonova, M.}, \bibinfo{author}{Egu{\'\i}luz, V.~M.} \&
  \bibinfo{author}{San~Miguel, M.}
\newblock \bibinfo{journal}{\bibinfo{title}{Noise in coevolving networks}}.
\newblock {\emph{\JournalTitle{Physical Review E}}}
  \textbf{\bibinfo{volume}{92}}, \bibinfo{pages}{032803}
  (\bibinfo{year}{2015}).

\bibitem{min2017fragmentation}
\bibinfo{author}{Min, B.} \& \bibinfo{author}{San~Miguel, M.}
\newblock \bibinfo{journal}{\bibinfo{title}{Fragmentation transitions in a
  coevolving nonlinear voter model}}.
\newblock {\emph{\JournalTitle{Scientific Reports}}}
  \textbf{\bibinfo{volume}{7}}, \bibinfo{pages}{12864} (\bibinfo{year}{2017}).

\bibitem{raducha2018coevolving}
\bibinfo{author}{Raducha, T.}, \bibinfo{author}{Min, B.} \&
  \bibinfo{author}{San~Miguel, M.}
\newblock \bibinfo{journal}{\bibinfo{title}{Coevolving nonlinear voter model
  with triadic closure}}.
\newblock {\emph{\JournalTitle{EPL (Europhysics Letters)}}}
  \textbf{\bibinfo{volume}{124}}, \bibinfo{pages}{30001}
  (\bibinfo{year}{2018}).

\bibitem{peralta2018analytical}
\bibinfo{author}{Peralta, A.~F.}, \bibinfo{author}{Carro, A.},
  \bibinfo{author}{San~Miguel, M.} \& \bibinfo{author}{Toral, R.}
\newblock \bibinfo{journal}{\bibinfo{title}{Analytical and numerical study of
  the non-linear noisy voter model on complex networks}}.
\newblock {\emph{\JournalTitle{Chaos: An Interdisciplinary Journal of Nonlinear
  Science}}} \textbf{\bibinfo{volume}{28}}, \bibinfo{pages}{075516}
  (\bibinfo{year}{2018}).

\bibitem{clauset2004finding}
\bibinfo{author}{Clauset, A.}, \bibinfo{author}{Newman, M.~E.} \&
  \bibinfo{author}{Moore, C.}
\newblock \bibinfo{journal}{\bibinfo{title}{Finding community structure in very
  large networks}}.
\newblock {\emph{\JournalTitle{Physical review E}}}
  \textbf{\bibinfo{volume}{70}}, \bibinfo{pages}{066111}
  (\bibinfo{year}{2004}).

\bibitem{girvan2002community}
\bibinfo{author}{Girvan, M.} \& \bibinfo{author}{Newman, M.~E.}
\newblock \bibinfo{journal}{\bibinfo{title}{Community structure in social and
  biological networks}}.
\newblock {\emph{\JournalTitle{Proceedings of the national academy of
  sciences}}} \textbf{\bibinfo{volume}{99}}, \bibinfo{pages}{7821--7826}
  (\bibinfo{year}{2002}).

\end{thebibliography}


\section*{Acknowledgements}

Financial  support  has  been  received  from  the  Agencia  Estatal  de  Investigacion  (AEI,  MCI,  Spain)  and Fondo Europeo de Desarrollo Regional (FEDER, UE), under Project PACSS (RTI2018-093732-B-C21/C22) and the Maria de Maeztu Program for units of Excellence in R\&D (MDM-2017-0711). T.R. would like to acknowledge the support from the National Science Centre
under Grant No. 2019/32/T/ST2/00133 and the support from the ZIP Programme
at the University of Warsaw. T.R. would like to thank Mateusz Wili\'nski
for useful suggestions in the early stage of this work
and Antonio Peralta for the advise on the analytical description.


\section*{Author contributions statement}

All authors conceived and designed the research,
T.R. conducted the simulations and performed analytical calculations,
all authors analyzed the results, wrote and reviewed the manuscript.


\section*{Additional information}

\textbf{Competing interests}
The authors declare no competing interests.


\end{document}